\documentclass[aps,prd,preprint,showpacs]{revtex4}
\usepackage{amssymb,amsmath,graphicx}

\begin{document}
\title{Influence on electron coherence from quantum electromagnetic fields in the presence of conducting plates}
\author{Jen-Tsung Hsiang}
\email{cosmology@mail.ndhu.edu.tw}
\author{Da-Shin Lee}
\email{dslee@mail.ndhu.edu.tw}\thanks{corresponding author.}
\affiliation{Department of Physics, National Dong Hwa University,
Hualien, Taiwan, R.O.C.}

\date{\today}

\begin{abstract}
The influence of electromagnetic vacuum fluctuations in the presence
of the perfectly  conducting plate on electrons is studied with an
interference experiment. The evolution of the reduced density matrix
of the electron is derived by the method of  influence functional.
We find that the plate boundary anisotropically modifies vacuum
fluctuations that in turn affect the electron coherence. The path
plane of the interference is chosen either parallel or normal to the
plate. In the vicinity of the plate, we show that the coherence
between electrons due to the boundary is enhanced in the parallel
configuration, but reduced in the normal case. The presence of the
second parallel plate is found to boost these effects.  The
potential relation between the amplitude change and phase shift of
interference fringes is pointed out. The finite conductivity effect
on electron coherence is discussed.
\end{abstract}

\maketitle

\section{Introduction}
Quantum coherence entails the existence of the interference effects
amongst alternative histories of the quantum states. These effects
are nevertheless not seen at the classical level. The suppression of
quantum coherence can be viewed as the result of the unavoidable coupling to the environment, and thus leads to the emergence of the classical behavior in terms of
incoherent mixtures. This environment-induced decoherence has been
studied with the idea of quantum open systems by coarse-graining
the environment where certain statistical measures are
introduced~\cite{ZU,HA,CAL1,CAL2,CAD}. Thereby, this averaged effect
appears as decoherence of the system of interest.

In modern cosmology, many efforts have been devoted to studying how
primordial perturbations, created quantum-mechanically during
inflation in the early universe, undergo the processes of
decoherence when their low momentum modes cross out the
horizon~\cite{CAL2,ST,RE}. They then reenter the horizon during the
radiation- or matter-dominated stage and thus act as the seeds of
temperature inhomogeneities in the cosmic microwave background as
well as the matter density inhomogeneities that lead to the
large-scale structure formation. In addition, special attention has
been paid to the possible observation of decoherence effects in
mesoscopic physics such as the phenomena of quantum tunneling, which
are affected by the coupling with a heat bath~\cite{CAD}. Recent
revival of interest in the decoherence phenomenon is motivated by
the study of the experimental realization of quantum computers in
which the central obstacle has proven to preventing the degradation
of the quantum coherence from the coupling of the computer to the
environment~\cite{NI}. Understanding of the aforementioned problems
relies on the deeper exploration of the decoherence dynamics driven
by the environment.

The quantum decoherence due to the interaction with the environment
has been discussed by considering the interference of the electron
states coupled to quantum electromagnetic fields in
vacuum~\cite{FO,MA}. It has been shown that the electron
interference pattern may be altered by particle creation and vacuum
fluctuations of electromagnetic fields, and the change might be
observed through the phase shift and the contrast change. However,
imposition of the boundary conditions on quantum fields may result
in the modification of vacuum fluctuations. The best-known example
is the attractive Casimir force between two parallel conducting
plates~\cite{CA}. This Casimir effect remains one of the least
intuitive consequences of quantum field theory~\cite{BA1,BA2,RO}.
Therefore, we expect that the presence of the boundary may further
influence the electron interference and gives rise to observable
effects. This type of the interference experiment can serve as a
probe to understanding the nature of quantum fluctuations~\cite{FO,
MA}.

Here we study the decoherence dynamics of the electron coupled to
quantum electromagnetic fields in the presence of the perfectly
conducting plate. We employ the closed-time-path formalism to
explore the evolution of the density matrix of the electron and
fields~\cite{SC}. In recent years, this nonequilibrium formalism has
been applied in particle physics and cosmology by one of
us~\cite{LEE}. The reduced density matrix of the electron can then
be derived with the method of influence functional, which takes
account of backreaction. We assume that the electron is initially in
a coherent superposition of two quantum states with their mean
trajectory along the distinct paths. Then the interference fringes
can be observed when these states are recombined. The phase shift
and amplitude reduction of the electron interference influenced by
quantum fields are obtained from the influence functional. The
leading effect of the decoherence functional comes from the
contribution evaluated along the prescribed electron's classical
trajectory defined by an applied potential. The validity of the
approximation will be discussed~\cite{FO,BR}. Note that this
coherence reduction is given by the double surface integrals of the
field strength correlation function defined in Minkowski spacetime
as we will see later. In this sense, it shares similar features with
the known Aharonov-Bohm effect where the phase shift of the electron
interference in the presence of the classical static magnetic field
depends on the magnetic flux in the region from which the electron
is absent. Here we instead consider the effects on the interference
from non-stationary quantum electromagnetic fields~\cite{HS}.

Our presentation is organized as follows. In Sec.~\ref{sec1}, we
introduce the closed-time-path formalism for describing the
evolution of the density matrix of a nonrelativistic electron
interacting with quantum electromagnetic fields. We then employ the
method of influence functional by tracing out the fields in the
Coulomb gauge in which we find the evolution of the reduced density
matrix for the electron with self-consistent backreaction. The
effect of decoherence can be realized by constructing the
decoherence functional from the influence functional under the
classical approximation in  Sec.~\ref{sec2}. In Sec.~\ref{sec3}, we
evaluate this decoherence functional for quantum electromagnetic
fields in the presence of the perfectly conducting plate and study
how coherence reduction of the electrons is affected by the modified
vacuum fluctuations due to the boundary. The finite conductivity
effect on electron coherence is discussed in Sec.~\ref{sec4}. The
results are summarized in Sec.~\ref{sec5}. In addition, in
App.~\ref{S:app1}, the nature of the gauge invariance in the
decoherence functional is considered by explicitly computing it with
an alternative gauge fixing. In App.~\ref{S:app2}, we outline the
method to convert a summation, which turns out to be slowly
convergent, into a rapidly convergent form.

The Lorentz-Heaviside units with $\hbar=c=1$ will be adopted unless
otherwise noted. The metric is
$\eta^{\mu\nu}=\text{diag}(1,-1,-1,-1)$.

\section{influence functional approach}\label{sec1}
We consider the dynamics of a nonrelativistic electron interacting
with quantum electromagnetic fields in the presence of the
conducting plate. In the Coulomb gauge, the electric and magnetic
fields can be expressed in terms of the vector potentials as:
\begin{equation}
    \mathbf{E}=-\nabla A_0-\dot{\mathbf{A}}_{\mathrm{T}}\,,\qquad\mathbf{B}=\nabla\times\mathbf{A}_{\mathrm{T}}\,,\label{E-B field}
\end{equation}
where $\mathbf{A}_{\mathrm{T}}$ is the transverse component of the
potential satisfying the  gauge condition,
$\nabla\cdot\mathbf{A}_{\mathrm{T}}=0$. The time-component of the
potential $A_0$ is not a dynamical field, but can be determined by
the Gauss law with the instantaneous Coulomb Green's function, which can be
defined by $\nabla^2G(\mathbf{x},\mathbf{y})=-\,\delta^3
(\mathbf{x}-\mathbf{y})$ subject to the boundary conditions. The
charge and current densities for a nonrelativistic electron may take
the form
\begin{equation}
    \varrho(x;\mathbf{q}(t))=e\;\delta^3(\mathbf{x}-\mathbf{q}(t))\,,\qquad\mathbf{j}_{\mathrm{T}}(x;\mathbf{q}(t))=e\;\dot{\mathbf{q}}(t)\,\delta^3(\mathbf{x}-\mathbf{q}(t))\,,\label{charge-current}
\end{equation}
with a coupling constant $e$. The current
$\mathbf{j}_{\mathrm{T}}$ satisfies the transverse condition
$\nabla\cdot\mathbf{j}_{\mathrm{T}}=0$. The Lagrangian of the electron-field
system is then given by the transverse components of the vector
potential as well as the coordinates $\mathbf{q}$ of the
non-relativistic electron,
\begin{equation}
    L[\mathbf{q},\mathbf{A}_{\mathrm{T}}]=\frac{1}{2}m\dot{\mathbf{q}}^2-V(\mathbf{q})-\frac{1}{2}\int\!d^3\mathbf{x}\,d^3\mathbf{y}\;\varrho(x;\mathbf{q})G(\mathbf{x},\mathbf{y})\varrho(y;\mathbf{q})+\int\!d^3\mathbf{x}\;\left[\frac{1}{2}(\partial_\mu\mathbf{A}_{\mathrm{T}})^2+\mathbf{j}_{\mathrm{T}}\cdot\mathbf{A}_{\mathrm{T}}\right]\,,\label{lagrangian}
\end{equation}
where an external potential $V$ is introduced so as to constrain the
motion of the electron to the prescribed path, and the Coulomb
electrostatic energy term is defined in the presence of the
boundary~\cite{BA2}.

The effect of electromagnetic fields on the electron interference
can be realized by the diagonal elements of the reduced density
matrix $\rho_r$, which is obtained by tracing out electromagnetic
fields in the density matrix of the electron and fields. Let us
consider that the initial density matrix at time $t_i$ can be
factorized as
\begin{equation}\label{initialcond}
    \rho(t_i)=\rho_{e}(t_i)\otimes\rho_{\mathbf{A}_{\mathrm{T}}}(t_i)\,
    ,
\end{equation}
and that initially the fields are assumed in thermal equilibrium at
temperature,  $\beta^{-1} $ with the density matrix
$\rho_{\mathbf{A}_{\mathrm{T}}}(t_i)$  given by
\begin{equation}\label{initialcondphi}
    \rho_{\mathbf{A}_{\mathrm{T}}}(t_i)=e^{-\beta H_{\mathbf{A}_{\mathrm{T}}}}\,,
\end{equation}
where $H_{\mathbf{A}_{\mathrm{T}}}$ is the Hamiltonian for the free
electromagnetic fields, constructed from Eq.~\eqref{lagrangian}.
Then the zero-temperature limit corresponding to the initial vacuum
state of the fields can be reached by taking $\beta \rightarrow
\infty$ limit. The electron-field system evolves unitarily according
to
\begin{equation}
    \rho(t_f)=U(t_f, t_i)\,\rho(t_i)\,U^{-1}(t_f,t_i )
\end{equation}
with $U(t_f,t_i)$ the time evolution operator. Thereafter, the final
state of the electron-field system in general becomes entangled due
to the interaction between them. The interaction between the
electron and fields will be assumed to be adiabatically switched on
in the remote past with $t_i\to-\infty$, and then switched off in
the remote future with $t_f\to\infty$. We then employ the
closed-time-path formalism to describe the evolution of the density
matrix of the electron-field. The reduced density matrix of the
electron, by tracing out the fields, becomes
\begin{eqnarray}
 \rho_r(\mathbf{q}_f,\tilde{\mathbf{q}}_f,t_f)&=&\int\!d\mathbf{A}_{\mathrm{T}}\;\bigl<\mathbf{q}_f,\mathbf{A}_{\mathrm{T}}\big|\rho(t_f)\big|\tilde{\mathbf{q}}_f,\mathbf{A}_{\mathrm{T}}\bigr>\nonumber \\
                   &=&\int\!d\mathbf{A}_{\mathrm{T}}\int\!d\mathbf{q}_1\,d\mathbf{A}_{1\mathrm{T}}\int\!d\mathbf{q}_2\,d\mathbf{A}_{2\mathrm{T}}\;\bigl<\mathbf{q},\mathbf{A}_{\mathrm{T}}\big|U(t_f,t_i)\big|\mathbf{q}_1,\mathbf{A}_{1\mathrm{T}}\bigr>\nonumber\\
                   &&\qquad\qquad\times\bigl<\mathbf{q}_1,\mathbf{A}_{1\mathrm{T}}\big|\rho(t_i)\big|\mathbf{q}_2,\mathbf{A}_{2,\mathrm{T}}\bigr>\bigl<\mathbf{q}_2,\mathbf{A}_{2\mathrm{T}}\big|U^{-1}(t_f,t_i)\big|\tilde{\mathbf{q}},\mathbf{A}_{\mathrm{T}}\bigr>\nonumber\\
                   &=&\int\!d\mathbf{q}_1\,d\mathbf{q}_2\int\!d\mathbf{A}_{\mathrm{T}}\,d\mathbf{A}_{1\mathrm{T}}\,d\mathbf{A}_{2\mathrm{T}}\int^{{\mathbf{q}}_f}_{\mathbf{q}_1}\!\mathcal{D}\mathbf{q}^+\int^{\tilde{\mathbf{q}}_f}_{\mathbf{q}_2}\!{\cal D}\mathbf{q}^-\int^{\mathbf{A}_{\mathrm{T}}}_{\mathbf{A}_{1\mathrm{T}}}\!\mathcal{D}\mathbf{A}_{\mathrm{T}}^+\int^{\mathbf{A}_{\mathrm{T}}}_{\mathbf{A}_{2\mathrm{T}}}\!\mathcal{D}\mathbf{A}_{\mathrm{T}}^-\nonumber\\
                   &&\qquad\qquad\times\int^{\mathbf{A}_{1\mathrm{T}}}_{\mathbf{A}_{2\mathrm{T}}}\!\mathcal{D}\mathbf{A}_{\mathrm{T}}^{\beta}\;\exp\biggl[i\int_{t_i}^{t_f}dt\;L[\mathbf{q}^+,\mathbf{A}_{\mathrm{T}}^+]-L[\mathbf{q}^-,\mathbf{A}_\mathrm{T}^-]\biggr]\nonumber\\
                   &&\qquad\qquad\qquad\qquad\qquad\times\exp\biggl[i\int_{t_i}^{t_i-i\beta}dt\;L_0[\mathbf{A}_{\mathrm{T}}^{\beta}]\biggr]\;\rho_{e}(\mathbf{q}_1,\mathbf{q}_2,t_i)\,.
\end{eqnarray}
Here we have introduced an identity in terms of a complete
set of eigenstates, $\big|\mathbf{q},\mathbf{A}_{\mathrm{T}}\bigr>$,
\begin{equation}
    \int\!d^3\mathbf{q}\,d\mathbf{A}_{\mathrm{T}}\;\big|\mathbf{q},\mathbf{A}_{\mathrm{T}}\bigr>\bigl<\mathbf{q},\mathbf{A}_{\mathrm{T}}\big|=1\,,
\end{equation}
with $\big|\mathbf{q},\mathbf{A}_{\mathrm{T}}\bigr>$ given by the
direct product of the states of the electron and those of
electromagnetic fields, namely,
$\big|\mathbf{q},\mathbf{A}_{\mathrm{T}}\bigr>=\big|\mathbf{q}\bigr>\otimes\big|\mathbf{A}_{\mathrm{T}}\bigr>$.
This identity has been inserted into the integrand so that the
matrix element of the time evolution operator can be expressed by
the path integral along  either the forward or backward time
evolution, represented by $\mathbf{q}^+$,
$\mathbf{A}_{\mathrm{T}}^+$, and $\mathbf{q}^-$,
$\mathbf{A}_{\mathrm{T}}^-$, respectively. The density matrix for
the thermal state of  fields corresponds to the evolution operator
of the fields $\mathbf{A}_{\mathrm{T}}^{\beta}$ along a path
parallel to the imaginary axis of complex time, and the time
arguments of the field operators are limited to the range between the complex time
$t_i$ and $t_i-i\beta$. Thus, the Green's functions of the vector
potentials possess the periodicity as the result of the cyclic
property of the trace as well as the bosonic nature of the field
operators.

Since the electron interacts with fields via a linear coupling, the
fields can be traced out exactly. Thus, we obtain the influence
functional for the electron by taking full account of the
backreaction. The physics becomes more transparent when we write the
evolution of the reduced density matrix in the following form
\begin{equation}
    \rho_r(\mathbf{q}_f,\tilde{\mathbf{q}}_f,t_f)=\int\!d^3\mathbf{q}_1\,d^3\mathbf{q}_2\;\mathcal{J}(\mathbf{q}_f,\tilde{\mathbf{q}}_f,t_f;\mathbf{q}_1,\mathbf{q}_2,t_i)\,\rho_{e}(\mathbf{q}_1,\mathbf{q}_2,t_i)\,,\label{evolveelectron}
\end{equation}
where the propagating function
$\mathcal{J}(\mathbf{q}_f,\tilde{\mathbf{q}}_f,t_f;\mathbf{q}_1,\mathbf{q}_2,t_i)$ is
\begin{equation}\label{propagator}
    \mathcal{J}(\mathbf{q}_f,\tilde{\mathbf{q}}_f,t_f;\mathbf{q}_1,\mathbf{q}_2,t_i)=\int^{\mathbf{q}_f}_{\mathbf{q}_1}\!\!\mathcal{D}\mathbf{q}^+\!\!\int^{\tilde{\mathbf{q}}_f}_{\mathbf{q}_2}\!\!\mathcal{D}\mathbf{q}^-\;\exp\left[i\int_{t_i}^{t_f}dt\left(L_{e}[\mathbf{q}^+]-L_{e}[\mathbf{q}^-]\right)\right]\mathcal{F}[\mathbf{j}^+_{\mathrm{T}},\mathbf{j}^-_{\mathrm{T}}]\,,
\end{equation}
and the electron Lagrangian $L_e[\mathbf{q}]$ is given by~\cite{BA2}
\begin{equation}\label{lag}
    L_e\bigl[\mathbf{q}\bigr]=\frac{1}{2}m\dot{\mathbf{q}}^2-V(\mathbf{q})-\frac{1}{2}\int\!d^3\mathbf{x}\,d^3\mathbf{y}\;\varrho(x;\mathbf{q})\,G(\mathbf{x},\mathbf{y})\,\varrho(y;\mathbf{q})\,.
\end{equation}
Here we introduce the influence functional $\mathcal{F}[\mathbf{
j}^+_{\mathrm{T}},\mathbf{j}^-_{\mathrm{T}}]$,
\begin{eqnarray}
    \mathcal{F}[\mathbf{j}^+_{\mathrm{T}},\mathbf{j}^-_{\mathrm{T}}]=\exp\biggl\{-\frac{1}{2}\,e^2\int d^4x\!\!\!&&\int\!d^4x'\Bigl[\;{\mathbf{j}^+_{\mathrm{T}}}_i(x;\mathbf{q}^+(t))\,\bigl<{\mathbf{A}^+_{\mathrm{T}}}^i(x){\mathbf{A}^+_{\mathrm{T}}}^j(x')\bigr>\,{\mathbf{j}^+_{\mathrm{T}}}_j(x';\mathbf{q}^-(t'))\Bigr.\biggr.\nonumber\\
    &&-{\mathbf{j}^+_{\mathrm{T}}}_i(x;\mathbf{q}^+(t))\,\bigl<{\mathbf{A}^+_{\mathrm{T}}}^i(x){\mathbf{A}^-_{\mathrm{T}}}^j(x')\bigr>\,{\mathbf{j}^-_{\mathrm{T}}}_j(x';\mathbf{q}^-(t'))\nonumber\\
    &&-{\mathbf{j}^-_{\mathrm{T}}}_i(x;\mathbf{q}^-(t))\,\bigl<{\mathbf{A}^-_{\mathrm{T}}}^i(x){\mathbf{A}^+_{\mathrm{T}}}^j(x')\bigr>\,{\mathbf{j}^+_{\mathrm{T}}}_j(x';\mathbf{q}^+(t'))\nonumber\\
    &&\biggl.\Bigl.+\,{{\bf j}^-_{{\rm T}}}_i(x;{\bf q}^-(t))\,\bigl<{\mathbf{A}^-_{\mathrm{T}}}^i(x){\mathbf{A}^-_{\mathrm{T}}}^j(x')\bigr>\,{\mathbf{j}^-_{\mathrm{T}}}_j(x';\mathbf{q}^-(t'))\Bigr]\biggr\}\,,\label{influencefun}
\end{eqnarray}
which contains full information about the influence of quantum
electromagnetic fields on the electron, and is a highly nonlocal
object. The Green's functions of the vector potential are defined by
\begin{eqnarray}
    \bigl<{\mathbf{A}^+_{\mathrm{T}}}^i(x){\mathbf{A}^+_{\mathrm{T}}}^j(x')\bigr>&=&\bigl<{\mathbf{A}_{\mathrm{T}}}^i(x){\mathbf{A}_{\mathrm{T}}}^j(x')\bigr>\,\theta(t-t')+\bigl<{\mathbf{A}_{\mathrm{T}}}^j(x'){\mathbf{A}_{\mathrm{T}}}^i(x)\bigr>\,\theta (t'-t)\,,\nonumber\\
    \bigl<{\mathbf{A}^-_{\mathrm{T}}}^i(x){\mathbf{A}^-_{\mathrm{T}}}^j(x')\bigr>&=&\bigl<{\mathbf{A}_{\mathrm{T}}}^j(x'){\mathbf{A}_{\mathrm{T}}}^i(x)\bigr>\,\theta(t-t')+\bigl<{\mathbf{A}_{\mathrm{T}}}^i(x){\mathbf{A}_{\mathrm{T}}}^j(x')\bigr> \,\theta (t'-t)\,,\nonumber\\
    \bigl<{\mathbf{A}^+_{\mathrm{T}}}^i(x){\mathbf{A}^-_{\mathrm{T}}}^j(x')\bigr>&=&\bigl<{\mathbf{A}_{\mathrm{T}}}^j(x'){\mathbf{A}_{\mathrm{T}}}^i(x)\bigr>\equiv\mathrm{Tr}\left\{\rho_{\mathbf{A}_{\mathrm{T}}}\,{\mathbf{A}_{\mathrm{T}}}^j(x'){\mathbf{A}_{\mathrm{T}}}^i(x)\right\}\,,\nonumber \\
    \bigl<{\mathbf{A}^-_{\mathrm{T}}}^i(x){\mathbf{A}^+_{\mathrm{T}}}^j(x')\bigr>&=&\bigr<{\mathbf{A}_{\mathrm{T}}}^i(x){\mathbf{A}_{\mathrm{T}}}^j(x')\bigr>\equiv\mathrm{Tr}\left\{\rho_{\mathbf{A}_{\mathrm{T}}}\,{\mathbf{A}_{\mathrm{T}}}^i(x){\mathbf{A}_{\mathrm{T}}}^j(x')\right\}\,,\label{noneqgreenfun}
\end{eqnarray}
and can be explicitly constructed as long as electromagnetic fields
are quantized subject to the boundary conditions. The retarded
Green's function and Hadamard function of  vector potentials are
defined respectively by
\begin{eqnarray}
    G_{R}^{ij}(x-x')&=&i\,\theta(t-t')\,\bigl<\left[{\mathbf{A}_{\mathrm{T}}}^i(x),{\mathbf{A}_{\mathrm{T}}}^j(x')\right]\bigr>\,,\label{commutator}\\
    G_{H}^{ij}(x-x')&=&\frac{1}{2}\,\bigl<\left\{{\mathbf{A}_{\mathrm{T}}}^i(x),{\mathbf{A}_{\mathrm{T}}}^j(x')\right\}\big>\,.\label{anticommutator}
\end{eqnarray}
Here the influence functional can be expressed in a more compact
form in terms of its phase and modulus by:
\begin{equation}
    \mathcal{F}[\,\mathbf{j}^+_{\mathrm{T}},\mathbf{j}^-_{\mathrm{T}}]=\exp\Big\{\mathcal{W}[\,\mathbf{j}^+_{\mathrm{T}},\mathbf{j}^-_{\mathrm{T}}]+i\,\Phi[\,\mathbf{j}^+_{\mathrm{T}},\mathbf{j}^-_{\mathrm{T}}]\Bigr\}\,,
\end{equation}
where
\begin{align}
    \Phi[\,\mathbf{j}^+_{\mathrm{T}},\mathbf{j}^-_{\mathrm{T}}]&=\frac{1}{2}\,e^2\!\!\int\!d^4x\!\!\int\!d^4x'\Bigl[\,{\mathbf{j}^+_{\mathrm{T}}}_i(x;\mathbf{q}^+)-{\mathbf{j}^-_{\mathrm{T}}}_i(x;\mathbf{q}^-)\Bigr]G_{R}^{ij}(x-x')\Bigl[\,{\mathbf{j}^+_{\mathrm{T}}}_j(x';\mathbf{q}^+)+{\mathbf{j}^-_{\mathrm{T}}}_j(x';\mathbf{q}^-)\Bigr]\,,\notag\\
    \mathcal{W}[\,\mathbf{j}^+_{\mathrm{T}},\mathbf{j}^-_{\mathrm{T}}]&=-\frac{1}{2}\,e^2\!\!\int\!d^4x\!\!\int\!d^4x'\Bigl[\,{\mathbf{j}^+_{\mathrm{T}}}_i(x;\mathbf{q}^+)-{\mathbf{j}^-_{\mathrm{T}}}_i(x;\mathbf{q}^-)\Bigr]G_{H}^{ij}(x-x')\Bigl[\,{\mathbf{j}^+_{\mathrm{T}}}_j(x';\mathbf{q}^+)-{\mathbf{j}^-_{\mathrm{T}}}_j(x';\mathbf{q}^-)\Bigr]\,.\label{phase-decoherence}
\end{align}
For a given initial state for the electron, the reduced density
matrix for the electron at time $t_f$ can be obtained from
Eq.~\eqref{evolveelectron} when the path integration over
$\mathbf{q}^{\pm}$ in Eq.~\eqref{propagator} is carried out.
Explicitly written out, the reduced density operator now becomes
\begin{eqnarray}
    \rho_r(\mathbf{q}_f,\tilde{\mathbf{q}}_f,t_f)&=&\int\!d^3\mathbf{q}_1\,d^3\mathbf{q}_2\left[\int^{\mathbf{q}_f}_{\mathbf{q}_1}\!\!\mathcal{D}\mathbf{q}^+\!\!\int^{\tilde{\mathbf{q}}_f}_{\mathbf{q}_2}\!\!\mathcal{D}\mathbf{q}^-\;\exp\biggl\{i\int_{t_i}^{t_f}dt\left(L_{e}[\mathbf{q}^+]-L_{e}[\mathbf{q}^-]\right)\biggr\}\right.\nonumber\\
            &&\qquad\qquad\left.\,\times\,\exp\biggl\{\mathcal{W}[\mathbf{q}^+,\mathbf{q}^-]\biggr\}\exp\biggl\{i\,\Phi[\mathbf{q}^+,\mathbf{q}^-]\biggr\}\right]\rho_{e}(\mathbf{q}_1,\mathbf{q}_2,t_i)\,.\label{evolve}
\end{eqnarray}
Let us now consider the initial electron state vector
$\bigl|\Psi(t_i)\bigr>$ to be a coherent superposition of two
localized states along worldlines $\mathcal{C}_1$ and
$\mathcal{C}_2$, respectively, after they leave the beam splitter at
the moment $t_i$,
\begin{equation}
    \bigl|\Psi(t_i)\bigr>=\bigl|\psi_1(t_i)\bigr>+\bigl|\psi_2(t_i)\bigr>\,.
\end{equation}
The density matrix of the electron state is then given by
\begin{eqnarray}
    \rho_e(t_i)&=&\bigl|\Psi(t_i)\bigr>\bigl<\Psi(t_i)\bigr|\\
                 &=&\rho_{11}(t_i)+\rho_{22}(t_i)+\rho_{21}(t_i)+\rho_{12}(t_i)\,,
\end{eqnarray}
where
$\rho_{mn}(t_i)=\bigl|\psi_m(t_i)\bigr>\bigl<\psi_n(t_i)\bigr|$. The
terms $\rho_{21}+\rho_{12}$ account for quantum interference,
because when the density matrix is realized in the coordinate basis,
we have
\begin{equation}
    \bigl<\mathbf{q}_i\big|\rho_e(t_i)\big|\mathbf{q}_i\bigr>=\left|\psi_1(\mathbf{q}_i,t_i)\right|^2+\left|\psi_2(\mathbf{q}_i,t_i)\right|^2+2\,\mathfrak{Re}\left\{\psi_2^*(\mathbf{q}_i,t_i)\psi_1^{\vphantom{*}}(\mathbf{q}_i,t_i)\right\}\,,
\end{equation}
which expresses the probability of finding an electron at
$(t_i,\mathbf{q}_i)$ in the superposed state. Therefore, at time
$t_f$, when the electron states are recombined at the location
$\mathbf{q}_f$,  the electron interference pattern can be described
by the diagonal elements of the reduced density matrix
$\bigl<\mathbf{q}_f\big|\rho_r(t_f)\big|\mathbf{q}_f\bigr>=\rho_r(\mathbf{q}_f,\mathbf{q}_f,t_f)$\,.

\section{decoherence functional in the classical approximation}\label{sec2}

The expression~\eqref{evolve} of the reduced density matrix at time
$t_f$ accounts for the full quantum effects of the electron, but the corresponding
path integral can not be carried out without
invoking further approximation~\cite{FO}. In general, the
interaction with quantum electromagnetic fields is expected to
perturb the electron's trajectory in a stochastic way about its mean
value, and to cause the electron wavefunction to spread~\cite{YU}.
It also fluctuates the phase of the wavefunction  such that the
phase coherence between electrons is lost.

Now considering the electron as a well-defined wave packet, its mean
trajectory follows the classical path constrained by an appropriate
external potential $V(\mathbf{q})$. The effect of the Coulomb
electrostatic attraction due to presence of the boundary is usually
small in the typical experiment configuration~\cite{FO}, and then
its influence on the trajectory can be ignored. In addition, the
backreaction from quantum field fluctuations, which is of the order
of the weak coupling $e^{2}/4\pi$ in the influence functional  also
has the ignorable correction to the classical paths as expected.
Furthermore, the finite spread of the wave packet of the electron
state, due to uncertainties on both position and momentum, can be
legitimately neglected as long as the electron's de Broglie
wavelength, $\lambda_{\mathrm{dB}}$ is much shorter than the
characteristic length scale associated with the accuracy of the
measurement $l$. Thus, as long as $l \gg\lambda_{\mathrm{dB}}$, the
wave packet can be viewed as it is sharply peaked in the electron's
position and momentum, and thus its quantum effects can be
ignored~\cite{FO}. As such, the leading effect of the decoherence
can be obtained by evaluating the propagating function
\eqref{propagator} along a prescribed classical path of the
electrons. Thereby, the diagonal components of the reduced density
matrix $\rho_r(\mathbf{q}_f,\mathbf{q}_f,t_f)$ now becomes
\begin{equation}
    \rho_r(\mathbf{q}_f,\mathbf{q}_f,t_f)=\bigl|\psi_1(\mathbf{q}_f,t_f)\bigr|^2+\bigl|\psi_2(\mathbf{q}_f,t_f)\bigr|^2+2\,e^{\mathcal{W}[\,\bar{\mathbf{j}}^1_{\mathrm{T}},\bar{\mathbf{j}}^2_{\mathrm{T}}]}\;\mathfrak{Re}\left\{e^{i\,\Phi[\,\bar{\mathbf{j}}^1_{\mathrm{T}},\bar{\mathbf{j}}^2_{\mathrm{T}}]}\psi_1^{\vphantom{*}}(\mathbf{q}_f,t_f)\,\psi_{2}^{*}(\mathbf{q}_f,t_f)\right\}\,,
\end{equation}
where the $\mathcal{W}$ and $\Phi$ functionals are evaluated along
the classical trajectories, $\mathcal{C}_1$ and $\mathcal{C}_2$. $\bar{\mathbf{j}}^{1,2}_{\mathrm{T}}$ is the classical current along the respective paths. The
evolution of the electron states $\psi_{1,2}(\mathbf{q}_f,t_f)$ is
governed by the Lagrangian $L_e$ in Eq.~\eqref{lag} due to the
ignorable backreaction effects.

The exponent of the modulus of the influence functional
$\mathcal{W}$, determined by the Hadamard function of vector
potentials, reveals decoherence between coherent electrons, while
its phase functional $\Phi$, related to the retarded Green's
function, results in an overall phase shift for the electron
interference pattern. Both effects arise from the interaction with
quantum fields. The decoherence functional can be obtained from the
expectation value of the anti-commutator of the vector potentials.
However, in the semiclassical Langevin equation to describe the
stochastic dynamics of the particle coupled to quantum fields, the
Hadamard function also determines the noise correlation function
from  quantum field fluctuations which cause the stochastic behavior
of the particle's trajectory~\cite{WU2}. Thus, we can conclude that
coherence reduction of the electrons is driven by field
fluctuations. On the other hand, the phase functional, which is
related to the retarded Green's function for the commutator of
vector potentials, link to the backreaction dissipation in the
Langevin equation on the dynamics of the particle~\cite{WU2}. Thus,
the phase shift may result from the backreaction dissipation from
quantum fields through particle creation that influences the mean
trajectory of the electrons. These two effects in the Langevin
equation obey the underlying fluctuation-dissipation theorem. In
this aspect,  the effects of quantum decoherence and the phase shift
are also likely related by the fluctuation-dissipation theorem.
However, very little work has been done to establish this relation.
It will be investigated in our future work. Here we only concentrate
on the effect of quantum decoherence induced by vacuum fluctuations
of electromagnetic fields.

In the classical approximation, with the help of
Eqs.~\eqref{charge-current} and \eqref{phase-decoherence}, the
decoherence factor, the $\mathcal{W}$ functional, can be expressed
as
\begin{equation}
    \mathcal{W}=-\frac{e^2}{2}\,\oint_Cdx_j\oint_Cdx'_k\,G_{H}^{jk}(x,x')\,,\label{w-fun}
\end{equation}
where $\mathbf{x},\mathbf{x}'\in C=C_1-C_2$ and $j,k=1,2,3$. The
curves $C_{1,2}$ are  the projection of the worldlines
$\mathcal{C}_{1,2}$ onto the hypersurface normal to the time axis in
Minkowski spacetime. Then, it is a straightforward calculation to
re-write the $\mathcal{W}$ functional in terms of the fields
$\mathbf{E}$ and $\mathbf{B}$ in a manifestly gauge invariant way,
\begin{equation}
    \mathcal{W}=-\frac{e^2}{8}\int_{\mathcal{C}}\!da_{\mu\nu}
    ^{\vphantom{'}}\int_{\mathcal{C}}\!da'_{\rho\sigma}\;\bigl<\left\{F^{\mu\nu}(x),F^{\rho\sigma}(x')\right\}\bigr>\,.\label{w-fun-E-B field}
\end{equation}
Apparently, the decoherence factor involves double surface integrals
of the expectation value of the anticommutator between the field
strength $F^{\mu\nu}$ as the area element $ d\sigma^{\mu\nu}$ of the
integral is bounded by a closed worldline of the electron
$\mathcal{C}$ in Minkowski spacetime. The closed worldline
$\mathcal{C}=\mathcal{C}_1-\mathcal{C}_2$ can be thought of as
moving electron along its path $C_1$ in the forward time direction
and then along the path $C_2$ in the backward time direction. By means of the
4-dimensional Stokes' theorem, we can write the $\mathcal{W}$
functional \eqref{w-fun-E-B field} as
\begin{equation}
\mathcal{W}=-\frac{e^2}{2}\oint_{\mathcal{C}}
\!dx^{\vphantom{'}}_{\mu}\,\oint_{\mathcal{C}}\!dx'_{\nu}\;G_{H}^{\mu\nu}(x,x')\,,
\label{W_gi}
\end{equation}
which involves the Hadamard function of the covariant vector
potentials.  It is consistent with the result in Ref.~\cite{FO}.

Note that although the expectation value of the vector potential
$\langle A^{\mu}(x)\rangle$ in the electromagnetic vacuum state
vanishes even in the presence of the boundary, the fluctuations of
fields are non-zero in general. The decoherence effect in
Eq.~\eqref{w-fun-E-B field} emerges as the result of the double
surface integrals of the non-vanishing field correlations in
Minkowski spacetime. Thus the decoherence is found sensitive to the
field strength in the region where the electron is excluded. In this
aspect, it may be regarded as the generalization to the
Aharonov-Bohm effect with time-independent classical electromagnetic
fields. In contrast, in our case, the decoherence effect is
essentially driven by  the non-static features of quantum fields.

\section{Evaluation of the $\mathcal{W}$ functional}\label{sec3}
\subsection{unbounded space}

As for illustration, let us start by considering  the
$\mathcal{W}_0$ functional for the unbounded space where
electromagnetic fields are initially  in the vacuum state~\cite{MA}.
The trajectory of the electrons can be dictated by an external
potential along the prescribed paths. The velocity of the electron
in the $x$ direction $v_x$ is assumed to be constant, while the
motion in the $z$ direction varies with time. Thus, the respective
worldlines of electrons are given by
$\mathcal{C}_{1,2}=(t,v_xt,0,\pm\zeta(t))$. The path function
$\zeta(t)$ is required to be sufficiently smooth to avoid enormous
photon production from the kinked corners and it may take the form,
\begin{equation}\label{path}
    \zeta(t)=R\,e^{-\frac{t^2}{T^2}}\,,
\end{equation}
where $2R$ is the effective path separation and $2T$ is the
effective flight time. The vector potential can be expressed by the
creation and annihilation operators as:
\begin{equation}\label{mode}
\mathbf{A}_{\mathrm{T}}(x)=\int\!\frac{d^3\mathbf{k}}{(2\pi)^{\frac{3}{2}}}\,
\frac{1}{\sqrt{2\omega}}\sum_{\lambda=1,2}\hat{\boldsymbol{\epsilon}}_{\lambda}(\mathbf{k})a_{\lambda}(\mathbf{k})\,e^{i\mathbf{k}\cdot\mathbf{x}-i\omega
t}+\text{H.C.}\,
\end{equation}
with $\omega=\left|\mathbf{k}\right|$. The polarization unit vectors
$\hat{\boldsymbol{\epsilon}}_{\lambda}$ obey the transversality
condition given by,
\begin{equation}
\sum_{\lambda=1,2}\hat{\boldsymbol{\epsilon}}^{i}_{\lambda}(\mathbf{k})\,\hat{\boldsymbol{\epsilon}}^{j}_{\lambda}(\mathbf{k})=\delta^{ij}-\frac{k^ik^j}{\left|\mathbf{k}\right|^2}\,.
\end{equation}
Since the $\mathcal{W}$ functional in Eq.~\eqref{W_gi} reveals
manifest Lorentz invariance, it proves more convenient to boost to a
frame $\mathfrak{S}$ moving with the velocity $u=(1,v_x,0,0)$ at
$y=z=0$, in which the electrons are seen to have transverse motion
in the $z$ direction only. Then, the $\mathcal{W}_0$
functional~\eqref{w-fun} can be obtained by a straightforward
calculation of the $z$--$z$ component of the vector potential
Hadamard function with the help of the mode expansion~\eqref{mode},
and reduces to
\begin{equation}
    \mathcal{W}_0=-2e^2\int\!\!\frac{d^3\mathbf{k}}{(2\pi)^3}\frac{1}{2\omega}\left[1-\frac{k_z^2}{\omega^2}\right]\left|\int\!\!dt\,\dot{\zeta}\cos\left(k_z\zeta\right)e^{i\omega t}\right|^2\,,
\end{equation}
where $\dot{\zeta}=d\zeta/dt$. We further simplify the calculation
by applying the dipole approximation,
$\cos\left(k_z\zeta\right)\simeq1$, consistent with the
non-relativistic limit. By using the path function \eqref{path}, the
decoherence functional $\mathcal{W}_0 $ ends up with
\begin{eqnarray}
    \mathcal{W}_0&\simeq&-2e^2\int\!\!\frac{d^3\mathbf{k}}{(2\pi)^3}\frac{1}{2\omega}\left[1-\frac{k_z^2}{\omega^2}\right]\left|\int\!dt\;\dot{\zeta}\;e^{i\omega t}\right|^2\nonumber\\
      &=&-\frac{e^2}{4\pi}R^2T^2\int_{-\infty}^{\infty}dk_z\int_{\left|k_z\right|}^{\infty}d\omega\left(\omega^2-k_z^2\right)e^{-\frac{1}{2}\omega^2T^2}\nonumber\\
       &=&-\frac{2e^2}{3\pi}\frac{R^2}{T^2} \left( \frac{1}{ c^2} \right) \,,
\end{eqnarray}
which is finite without the ultraviolet divergence. The absence of
the potential ultraviolet divergence can be seen from the
corresponding Fourier transform of the path function \eqref{path}
where the contribution from the high frequency modes with
$\omega\gtrsim\mathcal{O}(1/T)$ is exponentially suppressed. The result free of
ultraviolet divergence is quite general for the smooth path function
with the finite flight time.

In the nonrelativistic limit, since the transverse component of the
electron velocity $v_z$ is about $10^{-2} c$ in a typical
interference experiment, the decoherence factor $\mathcal{W}_0$,
proportional to $v^2_z$, will be of the order of $10^{-5}$ to
$10^{-6}$. Therefore, it is hardly to detect the loss of the
interference contrast due to vacuum fluctuations of quantum fields
in this unbounded case.

\subsection{presence of the single plate}\label{ss:single}

Now we consider the decoherence effect between the coherent
electrons under the influence of quantum electromagnetic fields in
the bounded region. In the presence of the perfectly conducting
plate, the tangential component of the electric field $\mathbf{E}$
as well as the normal component of the magnetic field $\mathbf{B}$
on the plate surface vanish. When the plate is placed at the $z=0$
plane, the boundary conditions of the fields $\mathbf{E}$ and
$\mathbf{B}$ on the plate give rise to
\begin{equation}
    A_0=0\,,\qquad\text{and}\qquad A_x=A_y=0\,,
\end{equation}
which lead to
\begin{equation}
    \frac{\partial{A}_z}{\partial z}=0
\end{equation}
as the result of the Coulomb gauge. The transverse vector potential
$\mathbf{A}_{\mathrm{T}}$ in the $z>0$ region is given
by~\cite{BA2},
\begin{eqnarray}
    \mathbf{A}_{\mathrm{T}}(x)&=&\int\!\frac{d^2\mathbf{k}_{\parallel}}{2\pi}\!\int_0^{\infty}\!\frac{dk_z}{(2\pi)^{1/2}}\;\frac{2}{\sqrt{2\omega}}\biggl\{a_1(\mathbf{k})\,\hat{\mathbf{k}}_{\parallel}\times\hat{\mathbf{z}}\,\sin k_zz\biggr.\nonumber \\
 &&\quad\quad\quad\biggl.+\;a_2(\mathbf{k})\left[i\,\hat{\mathbf{k}}_{\parallel}\left(\frac{k_z}{\omega}\right)\sin k_zz -\hat{\mathbf{z}}\left(\frac{k_{\parallel}}{\omega}\right)\cos k_zz\right]\biggr\}\;e^{i\mathbf{k}_{\parallel}\cdot\mathbf{x}_{\parallel}-i\omega t}+\text{H.C.}\,,
\end{eqnarray}
where the circumflex identifies unit vectors. The position vector
$\mathbf{x}$ is denoted by $\mathbf{x}=( \mathbf{x}_{\parallel},z)$
where $\mathbf{x}_{\parallel}$ is the components parallel to the
plate. Similarly, the wave vector is expressed by  $\mathbf{k}=(
\mathbf{k}_{\parallel},k_ z)$ with $\omega^2=k_{\parallel}^2+k_z^2$.
The creation and annihilation operators obey the commutation
relations
\begin{equation}
    [a_{{\lambda}^{\vphantom{l}}}^{\vphantom{\dagger}}(\mathbf{k}),a_{\lambda'}^{\dagger}(\mathbf{k}')]=\delta_{{\lambda}{\lambda}'}\,\delta(\mathbf{k}_{\parallel}^{\vphantom{'}}-\mathbf{k}'_{\parallel})\,\delta(k_z^{\vphantom{'}}-k'_z)\,,
\end{equation}
and otherwise are zero.

\begin{figure}
\centering
    \scalebox{0.7}{\includegraphics{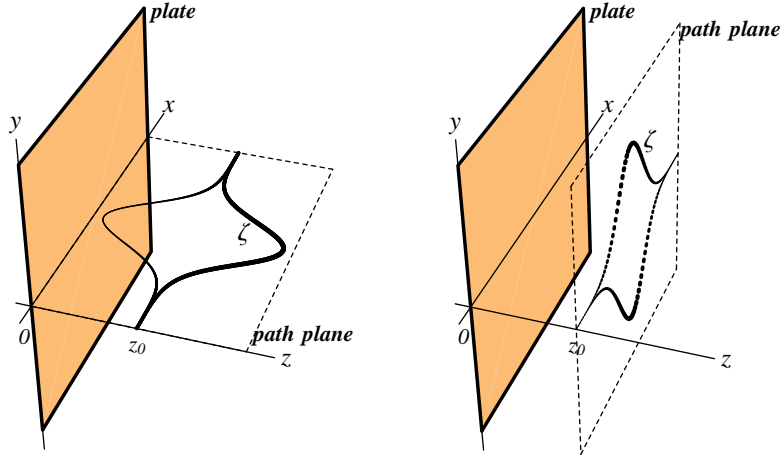}}
    \caption{Two different orientations of the electron path plane relative to the conducting plate.}\label{Fi:pp}
\end{figure}
\begin{figure}
\centering
    \scalebox{1.0}{\includegraphics{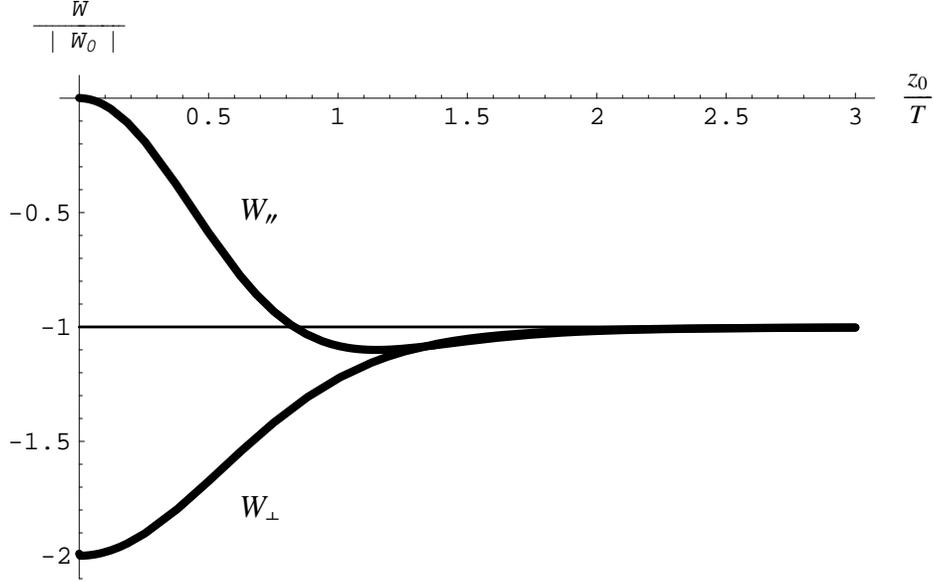}}
    \caption{The decoherence functional $\mathcal{W}$ for the single-plate boundary as a function of the ratio $\xi=z_0/T$.}\label{Fi:paraperp1}
\end{figure}

The path plane on which the electrons travel can be either parallel
or perpendicular to the plate. When the path plane is normal to the
conducting plate as shown in Fig.~\ref{Fi:pp}, the electron
worldlines are given by
$\mathcal{C}_{1,2}=(t,v_xt,0,z_0\pm\zeta(t))$. We will choose a
frame $\mathfrak{S}$ which moves along the worldline $(t,v_xt,0,z_0)$ and has the
same orientation as the laboratory frame. In this frame, the
electrons are seen to have sideways motion in the $z$ direction
only. Then the $\mathcal{W}_{\perp}$ functional depends on the
$z$--$z$ component of the vector potential Hadamard function, which
is given by
\begin{equation}
    G_{H}^{zz}(x,x')=\frac{1}{2}\int\frac{d^3\bf{k}}{(2\pi)^3}\frac{1}{2\omega}\left(\frac{k_{\parallel}}
    {\omega}\right)^2\cos k_zz\,\cos k_zz'\;e^{i\mathbf{k}_{\parallel}^{\vphantom{'}}\cdot(\mathbf{x}_{\parallel}^{\vphantom{'}}
    -\mathbf{x}'_{\parallel})-i\omega(t-t')}+\text{C.C.}\,.
    \label{HD-fun_zz}
\end{equation}
Thus, the decoherence functional $\mathcal{W}_{\perp}$ can be obtained as:
\begin{eqnarray}
    \mathcal{W}_{\perp}&=&-\frac{e^2}{2}\left(\int_{C_1}dz\int_{C_1}dz'+\int_{C_2}dz\int_{C_2}dz'-\int_{C_1} dz\int_{C_2}dz'-\int_{C_2}dz\int_{C_1}dz'\right)G_{H}^{zz}(x,x')\nonumber \\
  &=&-\frac{e^2}{4}\int\!dt\,\dot{\zeta}\;dt'\,\dot{\zeta}'\!\int\frac{d^3\bf{k}}{(2\pi)^3}\;\frac{1}{2\omega}\left(\frac{k_{\parallel}}
    {\omega}\right)^2\biggl\{\cos k_z(z_0+\zeta)\cos k_z(z_0+\zeta')\biggr.\nonumber \\
   &&\qquad\qquad\qquad\qquad+\cos k_z(z_0-\zeta)\cos k_z(z_0-\zeta')+\cos k_z(z_0-\zeta)\cos k_z(z_0+\zeta')\nonumber \\
   &&\biggl.\qquad\qquad\qquad\qquad+\cos k_z(z_0+\zeta)\cos k_z(z_0-\zeta')\biggr\}e^{-i\omega(t-t')}+\text{C.C.}\nonumber\\
   &=&-2e^2 \int\!\frac{d^3\bf{k}}{(2\pi)^3}\;\frac{1}{2\omega}\left[1-\frac{k_z^2}{\omega^2}\right]\biggl[1+e^{i2k_zz_0}\biggr]\left|\int\!dt\;\dot{\zeta}\cos k_z\zeta\;e^{-i\omega t}\right|^2 \,.
   \label{W_perp_single}
  \end{eqnarray}
Then, under the dipole approximation, we arrive at:
\begin{equation}
    \mathcal{W}_{\perp}=\mathcal{W}_0\left\{1+ \frac{3}{32\xi^3}\left[-4\xi+\sqrt{2\pi}\left(1+4\xi^2\right)e^{-2\xi^2}\operatorname{Erfi}(\sqrt{2}\xi)\right]\right\}\label{W_perp}
\end{equation}
with the path function given by Eq.~\eqref{path}. Here the
corrections to the decoherence functional due to the presence of the
conducting plate is expressed in terms of the ratio of the effective
distance of the electrons to the plate over the parameter $T$, i.e.,
$\xi=z_0/T$. The imaginary error function $\operatorname{Erfi}(z)$
is defined by
\begin{equation}
\operatorname{Erfi}(z)\equiv-i\operatorname{Erf}(i\,z)=\frac{2}{\sqrt{\pi}}\int^z_0\!ds\,e^{s^2}\,.
\end{equation}
Asymptotically, the ratio
$\mathcal{W}_{\perp}/\left|\mathcal{W}_0\right|$ is given by
\begin{equation}\label{ds}
    \frac{\mathcal{W}_{\perp}}{\left|\mathcal{W}_0\right|}=\begin{cases}
                                                                -2+\displaystyle\frac{8}{5}\xi^2+\mathcal{O}(\xi^4)\,, &\xi\to0\,;\\
                                                                &\\
                                                                -1-\displaystyle\frac{3}{16}\frac{1}{\xi^4}+\mathcal{O}(\frac{1}{\xi^{6}})\,, &\xi\to\infty\,.
                                                           \end{cases}
\end{equation}
As shown in Fig.~\ref{Fi:paraperp1}, the effects of coherence
reduction by vacuum fluctuations in the presence of the boundary are
strikingly deviated from that without the boundary. It can be
understood by the fact that the presence of the perfectly conducting
plate modifies zero-point fluctuations of the fields which manifest
themselves so as to influence the dynamics of decoherence in the
electron interference.

In particular, when the path plane lies normal to the plate, we find
that the modified vacuum fluctuations due to the boundary further
reduce the electron coherence, then in turn suppress the contrast of
the interference fringes for all values of $\xi$. It is found that
for small $\xi$,
$\mathcal{W}_{\perp}\approx2\mathcal{W}_0$~\cite{MA}. To understand
this, here we provide an explanation in contrast to the fictitious
dipole interpretation suggested by Ref.~\cite{MA}. Let us note that,
in the reference frame $\mathfrak{S}$, the relevant component of the
electromagnetic fields in this case is the $\mathbf{E}_z$ field,
which is perpendicular to the conducting plate. The effect of the
neutral conducting plate can be achieved by placing an image charge
at the location symmetrical to the original charge with respect to
the plate. The image charge shall carry the opposite sign to the
real one as required by the boundary conditions. As such, the
$\mathbf{E}_z$ field produced by the image charge is almost the same
as that by the original one so as to make the total $\mathbf{E}_z$
field near the surface twice that in the unbounded case. Thus, the
decoherence effect is doubled~\cite{MA}. However, when the ratio
$\xi$ increases, the suppression of electron coherence is alleviated
as expected and finally reduces to the result without the boundary
in the limit $\xi\rightarrow\infty$. Also note that the ratio $\xi$
can not infinitesimally go to zero because $z_0$ has to be larger
than $R$ to constrain the electrons on one side of the plate.

On the other hand, when the path plane lies parallel to the
conducting plate, here the electron worldlines are given by
$\mathcal{C}_{1,2}=(t,v_xt,\pm\zeta(t),z_0)$.  The same reference
frame $\mathfrak{S}$ is chosen so that the electrons are seen to
move in the $y$ direction. Then, the $y$--$y$ component of the
vector potential Hadamard function becomes relevant to the
$\mathcal{W}_{\parallel}$ and it is given by,
\begin{equation}
    G_{H}^{yy}(x,x')=\frac{1}{2}\int\frac{d^3\mathbf{k}}{(2\pi)^3}\frac{1}{2\omega}\left\{\sin^2\chi+\frac{k_z^2}{\omega^2}\cos^2\chi\right\}\sin k_zz\,\sin k_zz'\;e^{i\mathbf{k}_{\parallel}^{\vphantom{'}}\cdot(\mathbf{x}_{\parallel}^{\vphantom{'}}-\mathbf{x}'_{\parallel})-i\omega(t-t')}+\text{C.C.}\,,
\end{equation}
where $\chi$ is the angle between $\hat{\mathbf{y}}$ and
$\hat{\mathbf{k}}_{\parallel}$. We then obtain the decoherence
functional $\mathcal{W}_{\parallel}$,
\begin{eqnarray}
    \mathcal{W}_{\parallel}&=&-e^2\left(\int_{C_1}dy\int_{C_1}dy'-\int_{C_1}dy\int_{C_2}dy'\right)G_{H}^{yy}(x,x')\nonumber\\
                              &=&-2e^2\int\!\frac{d^3\mathbf{k}}{(2\pi)^3}\;\frac{1}{2\omega}\left[1-\frac{k_y^2}{\omega^2}\right]\biggl[1-e^{i2k_zz_0}\biggr]\left|\int\!dt\;\dot{\zeta}\cos k_z\zeta\;e^{-i\omega t}\right|^2\,.\label{W-parallel-single}
\end{eqnarray}
Following the same approximation to obtain Eq.~\eqref{W_perp}, the
$\mathcal{W}_{\parallel}$ now is given by
\begin{equation}
    \mathcal{W}_{\parallel}=\mathcal{W}_0\left\{1-\frac{3}{64\xi^3}\left[4\xi\left(1+4\xi^2\right)-\sqrt{2\pi}\left(1+16\xi^4\right)e^{-2\xi^2}\operatorname{Erfi}(\sqrt{2}\xi)\right]\right\}\,,
\end{equation}
and asymptotically $\mathcal{W}_{\parallel}/\left|\mathcal{W}_0\right|$ is obtained as
\begin{equation}
    \frac{\mathcal{W}_{\parallel}}{\left|\mathcal{W}_0\right|}=\begin{cases}
                                                                -\displaystyle\frac{16}{5}\xi^2+\frac{144}{35}\xi^4+\mathcal{O}(\xi^6)\,, &\xi\to0\,;\\
                                                                &\\
                                                                -1-\displaystyle\frac{3}{16}\frac{1}{\xi^4}+\mathcal{O}(\frac{1}{\xi^6})\,, &\xi\to\infty\,.
                                                           \end{cases}
\end{equation}

In contrast to the perpendicular case, near the plate surface where
$ \xi\ll1$, the electron coherence is enhanced. The loss of
coherence originally due to vacuum fluctuations in the unbounded
space is almost completely compensated by the induced fluctuations
due to the boundary, especially in the limit of $\xi\to0$. For small
$\xi$, we find that $\mathcal{W}_{\parallel}\approx0$. Apparently,
in the reference frame $\mathfrak{S}$, the $\mathbf{E}_y$ component,
which is parallel to the plate, is crucial. As required by the
boundary conditions, the presence of the image charge renders the
$\mathbf{E}_y$ field almost zero near the plate surface, leading to
the vanishing field fluctuations. Thus, it is not  so surprising
that the electron coherence is restored near the plate surface.
However, when the ratio $\xi$ is much greater than unity, we expect
that the orientation of the path plane becomes irrelevant. The
influence of the boundary on electron coherence is negligible. The
decoherence effect reduces to the result in the perpendicular
configuration, and then to that in the unbounded case in the limit
$\xi \rightarrow \infty$. We can see from Fig.~\ref{Fi:paraperp1}
that the presence of the boundary makes the electrons more coherent
for small $\xi$, but less coherent for large $\xi$ in the parallel
configuration.

The presence of the conducting plate anisotropically modifies the
electromagnetic vacuum fluctuations that in turn influence the
dynamics of the electrons coupled to the fields. In Ref.~\cite{YU},
the authors investigate the Brownian motion of the test particle
coupled to quantized electromagnetic fields. An anisotropical
modification in the mean squared fluctuations of the velocity near
the conducting plate is found. Since the mean squared fluctuations
of the velocity reflect vacuum fluctuations of fields, it is
concluded that close to the plate, the electromagnetic vacuum
fluctuations are suppressed in the direction transverse to the
plate, compared to the unbounded case, while fluctuations are
enhanced in the longitudinal direction. This is consistent with our
results.

\subsection{presence of the double plates}
In the presence of double plates, we place the second plate at
$z=a$, in addition to the one at $z=0$. Thus, the transverse vector
potential $\mathbf{A}_{\mathrm{T}}$ for the bounded region between
the $z=0$ and $z=a$ planes can be expressed by~\cite{BA1}
\begin{align}
    \mathbf{A}_{\mathrm{T}}(x)&=\sqrt{\frac{2}{a}}\;{\sum_{n=0}^{\infty}}''\!\int\!\frac{d^2\mathbf{k}_{\parallel}}{2\pi}\frac{1}
    {\sqrt{2\omega_n}}\biggl\{a_1(\mathbf{k}_{\parallel},n)\,\hat{\mathbf{k}}_{\parallel}\times\hat{\mathbf{z}}\,\sin\frac{n\pi}{a}z\biggr.\notag\\
    &\qquad\biggl.+\;a_2(\mathbf{k}_{\parallel},n)\left[i\,\hat{\mathbf{k}}_{\parallel}\left(\frac{n\pi}{\omega_na}\right)\sin\frac{n\pi}{a}z-\hat{\mathbf{z}}\left(\frac{k_{\parallel}}{\omega_n}\right)\cos\frac{n\pi}{a}z\right]\biggr\}e^{i\mathbf{k}_{\parallel}\cdot\mathbf{x}_{\parallel}-i\omega_nt}+\text{H.C.}\,.
\end{align}
The double prime on $\sum$ assigns an extra normalization factor
$1/\sqrt{2}$ to the $n=0$ mode. The discrete frequencies $\omega_n$
of the allowed modes for the double-plate  boundary are
\begin{equation}
    \omega_n^2=k_{\parallel}^2+\left(\frac{n\pi}{a}\right)^2\,. \label{omega_dp}
\end{equation}
Moreover, the creation and annihilation operators obey the commutation
relations
\begin{equation}
    [a_{{\lambda}^{\vphantom{l}}}^{\vphantom{\dagger}}(\mathbf{k}_{\parallel}^{\vphantom{'}},n),a_{{\lambda}'}^{\dagger}(\mathbf{k}'_{\parallel},n')]=\delta_{{\lambda}{\lambda}'}\,\delta_{nn'}\,\delta(\mathbf{k}^{\vphantom{'}}_{\parallel}-\mathbf{k}'_{\parallel})\,,
\end{equation}
and otherwise vanish.

As in the single plate case, we consider that the path plane lies
either parallel or perpendicular to the plates. In the perpendicular
case, we assume that the electrons move along their worldlines,
described by $\mathcal{C}_{1,2}=(t,v_xt,0,\frac{a}{2}\pm\zeta(t))$,
where the path function $\zeta$ is given by Eq.~\eqref{path}. As
before, we choose the frame $\mathfrak{S}$ with $z_0=a/2$, in which
the electrons are observed to move in the $z$ direction. Thereby, the relevant
component of the vector potential Hadamard function is the
$z$--$z$ component,
\begin{equation}
    G_{H}^{zz}(x,x')=\frac{1}{2a}\sum_{n=-\infty}^{\infty}\int\!\frac{d^2\mathbf{k}_{\parallel}}{(2\pi)^2}\frac{1}{2\omega_n}\left(\frac{k_{\parallel}}{\omega_n}\right)^2\!\!\cos\frac{n\pi}{a}z\,\cos\frac{n\pi}{a}z'\;e^{i\mathbf{k}_{\parallel}^{\vphantom{'}}\cdot(\mathbf{x}_{\parallel}^{\vphantom{'}}-\mathbf{x}'_{\parallel})-i\omega_n(t-t')}+\text{C.C.}\,.
\end{equation}
Thus, the decoherence functional $\mathcal{W}_{\perp}$ now becomes
\begin{equation}
   \mathcal{W}_{\perp}=-\frac{2e^2}{a}\sum_{n=\text{even}}\!\int\!\frac{d^2\bf{k}_{\parallel}}{(2\pi)^2}\frac{1}{2\omega_n}
   \left[1-\frac{n^2\pi^2}{\omega_n^2a^2}\right]\left|\int\!dt\;\dot{\zeta}\cos\frac{n\pi}{a}\zeta\;e^{-i\omega_nt}\right|^2\,.
   \label{W-normal-double}
\end{equation}
Then, by applying the dipole approximation, it reduces  to
\begin{eqnarray}
    \mathcal{W}_{\perp}&\simeq& -\frac{e^2}{2}\,\frac{R^2}{T\,a}\sum_{n=-\infty}^{\infty}\left[\left|n\right|\varsigma\,e^{-\frac{1}{2}\,n^2\varsigma^2}+\sqrt{\frac{\pi}{2}}\left(1-n^2\varsigma^2\right)\operatorname{Erfc}(\frac{\left|n\right| \varsigma}{\sqrt{2}}) \right]\,,\label{W-normal-double-approx}
\end{eqnarray}
with $\varsigma=2\pi T/a $.  The complementary error function,
$\operatorname{Erfc}(z)$, is defined as
\begin{equation}
    \operatorname{Erfc}(z)\equiv1-\operatorname{Erf}(z)=\frac{2}{\sqrt{\pi}}\int^{\infty}_z\!ds\;e^{-s^2}\,.\label{erfc}
\end{equation}

Let us now consider the limit $\varsigma\gg1$, that is, $a\ll T$,
where the plate separation is much smaller than the flight time. In
this limit, the decoherence functional
\eqref{W-normal-double-approx} is dominated by the $n=0$ term with
$\operatorname{Erfc}(0)=1$, while the $n\ne0$ terms are
exponentially suppressed due to large $\varsigma$. Hence, the
$\mathcal{W}_{\perp}$ functional can be approximated by
\begin{eqnarray}
    \mathcal{W}_{\perp}&\simeq&-\frac{e^2}{2}\,\frac{R^2}{T\,a}\left\{\sqrt{\frac{\pi}{2}}+\frac{2}{\pi}\left(\frac{a}{T}\right)\,e^{-2\pi^2\frac{T^2}{a^2}}+\cdots\right\}\,.
\end{eqnarray}
It can be seen that the result is very small for $R<a\ll T$.
However, compared with the unbounded case, the ratio
$\mathcal{W}_{\perp}/\left|\mathcal{W}_0\right|$ is
\begin{equation}
    \frac{\mathcal{W}_{\perp}}{\left|\mathcal{W}_0\right|}=-\frac{3 \pi^{\frac{3}{2}}}{4 \sqrt{2}}\,\frac{T}{a}\ll-1\,,
\end{equation}
thus more significantly degrading electron coherence. Nonetheless,
the ratio $a/T$ can not indefinitely go to zero, and is bounded by
$2R/T$ from below since the plate separation can not be smaller than
the path separation.

As the ratio $a/T$ becomes much greater than unity, the value of the
decoherence functional  reduces to the unbounded case. To see it, we
convert Eq.~\eqref{W-normal-double} to a form suited for this limit
with the method outlined in App.~\ref{S:app2}. The
$\mathcal{W}_{\perp}$ functional now takes the form
\begin{eqnarray}
    \mathcal{W}_{\perp}&=&-\frac{e^2}{2}\,\frac{R^2}{T\,a}\left\{\frac{8}{3\varsigma}
      +4\sum_{n=1}^{\infty}\left[-\frac{\varsigma}{2\pi^2}\frac{1}{n^2}+\frac{1}{(2\pi)^{\frac{1}{2}}}\frac{1}{n^3}\left(n^2+\frac{\varsigma^2}{4\pi^2}\right)e^{-2\left(\frac{n\pi}{\varsigma}\right)^2}\operatorname{Erfi}(\sqrt{2}\,\frac{n\pi}{\varsigma})\right]\right\}\nonumber\\
                        &\simeq& -e^2\,\frac{R^2}{T\,a}\left\{\frac{4}{3\varsigma}+\frac{\varsigma^3}{2\pi^4}\sum_{n=1}^{\infty}\frac{1}{n^4}\,e^{-2\left(\frac{\pi}{\varsigma}\right)^2n^2}+\cdots\right\}\,,\qquad\text{for $\varsigma\ll 1$}\,.
\end{eqnarray}
Here the last line is obtained by Taylor-expanding the terms in the
square bracket. Thus, we have, in the limit $a\gg T$,
\begin{equation}\label{pq}
   \mathcal{W}_{\perp}\simeq\mathcal{W}_0\left\{1+6\,\frac{T^4}{a^4}\,e^{-\frac{a^2}{2T^2}}+\cdots\right\}\,.
\end{equation}
The first term is the contribution to the decoherence effect from
vacuum fluctuations without the  boundary, while the second term,
although exponentially small, is the correction due to the presence
of the double plates.

\begin{figure}
\centering
    \scalebox{1.0}{\includegraphics{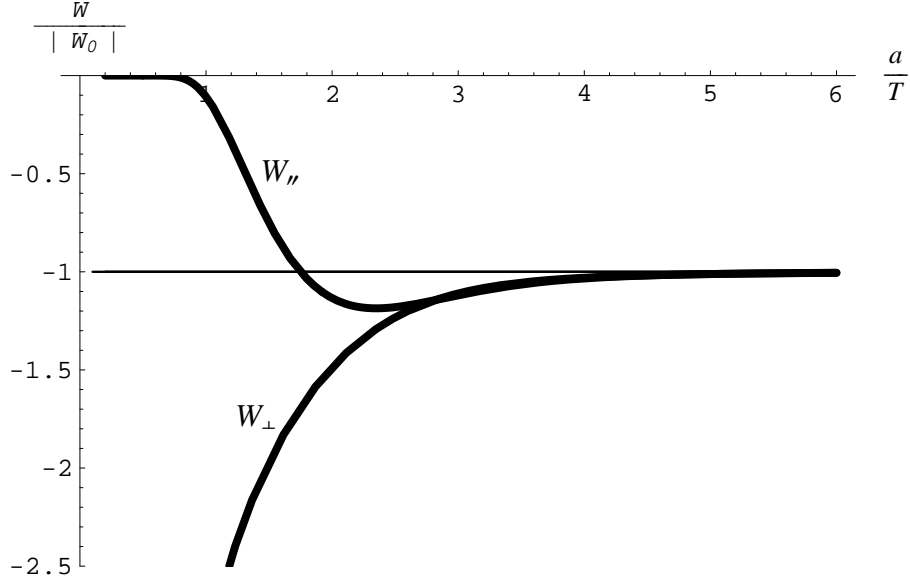}}
    \caption{The decoherence functional $\mathcal{W}$ for the double-plate boundary
    as a function of the ratio $a/T$.}\label{Fi:paraperp2}
\end{figure}

In Fig.~\ref{Fi:paraperp2}, the ratio of the $\mathcal{W}_{\perp}$
functional over the absolute value of $\mathcal{W}_0$ is plotted for
a very wide range of $a/T$. It is shown that vacuum fluctuations
arising from the presence of the plates always degrades electron
coherence for the perpendicular case as expected from the single
plate case. In addition, introducing the second plate seems to boost
fluctuations so as to further reduce the electron coherence
significantly in the limit of $a/T\ll1$, where the effect of the
boundary becomes important.

In the parallel case, the electron worldlines are given by
$\mathcal{C}_{1,2}=(t,v_xt,\pm\zeta,\frac{a}{2})$, and the same
reference frame $\mathfrak{S}$ is chosen. Then in this frame the electrons are
observed to move in the $y$ direction only. The contributing
component of the vector potential Hadamard function is the $y$--$y$
component, given by:
\begin{eqnarray}
G_{H}^{yy}(x,x')&=&\frac{1}{2a}\sum_{n=-\infty}^{\infty}\int\!\frac{d^2\mathbf{k}_{\parallel}}{(2\pi)^2}\frac{1}{2\omega_n}\biggl\{\sin^2\chi\,\sin\frac{n\pi}{a}z\,\sin\frac{n\pi}{a}z'\biggr.\nonumber\\
   &&\biggl.+\left(\frac{n\pi}{\omega_na}\right)^2\cos^2\chi\,\sin\frac{n\pi}{a}z\,\sin\frac{n\pi}{a}z'\biggr\}e^{i\mathbf{k}_{\parallel}^{\vphantom{'}}\cdot(\mathbf{x}_{\parallel}^{\vphantom{'}}-\mathbf{x}'_{\parallel})-i\omega_n(t-t')}+\text{C.C.}\,,
\end{eqnarray}
where $\chi$ is the angle between $\hat{\mathbf{y}}$ and
$\hat{\mathbf{k}}_{\parallel}$. Then, the decoherence functional
$\mathcal{W}_{\parallel}$ is
\begin{equation}\label{wew}
    \mathcal{W}_{\parallel}=-\frac{2e^2}{a}\sum_{n=\text{odd}}\!\int\!\frac{d^2\mathbf{k}_{\parallel}}{(2\pi)^2}\frac{1}{2\omega_n}\left[1-\frac{k_y^2}{\omega_n^2}\right]\left|\int\!dt\;\dot{\zeta}\cos k_y\,\zeta\;e^{-i\omega_nt}\right|^2\,.
\end{equation}
Therefore, we obtain
\begin{equation}
    \mathcal{W}_{\parallel}\simeq-\frac{e^2}{2}\,\frac{R^2}{T\,a}\sum_{n=1}^{\infty}\left[q_n\,\varsigma\,e^{-\frac{1}{2}q_n^2\varsigma^2}+\sqrt{\frac{\pi}{2}}\left(1+q_n^2\varsigma^2\right)\operatorname{Erfc}(\frac{q_n\varsigma}{\sqrt{2}})\right]\,,\label{W-parallel-double-approx}\end{equation}with $q_n=n-1/2$. The same approximation we invoked in the
perpendicular case can be applied  here. The result of
Eq.~\eqref{W-parallel-double-approx} is shown in
Fig.~\ref{Fi:paraperp2}, which reveals the similar features as in
the single-plate case.

In the limit $\varsigma\gg1$ or $T\gg a$, the dominant contribution
to Eq.~\eqref{W-parallel-double-approx} comes from the $n=1$ term,
and the decoherence functional can be further approximated by
\begin{equation}
    \mathcal{W}_{\parallel} \simeq -\pi\,e^2\,\frac{R^2}{a^2}\,e^{-\frac{\pi^2}{2}\frac{T^2}{a^2}}\,,\label{we}
\end{equation}
which is exponentially small as $a/T\rightarrow 0$. It can be
interpreted as the fact that the presence of the double-plate
boundary may further suppress vacuum fluctuations in the direction
parallel to the conducting plates as compared with the single-plate
case, and thus enhances the electron coherence. An interesting
feature of the double-plate case can be seen from
Fig.~\ref{Fi:paraperp2} that the plot has a rather wide plateau for
the small $a/T$ up to the value $a/T\sim 1$ within which no
appreciable loss of electron coherence could be observed. It can be
understood by the fact that when both plates come close to one
another, the dominant contribution to Eq.~\eqref{wew} comes from the
$n=1$ modes. Since their frequencies, $\omega_1\geq\pi/a$ for all
${\bf k}_{\parallel}$ obtained from Eq.~\eqref{omega_dp}, have
become sufficiently high due to small $a$, the contributions of
those modes are exponentially suppressed as can be seen from the
absolute value of the integral in Eq.~\eqref{wew}. This is quite
different from the single-plate case.

Next, in the other limit $\varsigma\ll1$ or $T\ll a$, it is
straightforward to show that the $\mathcal{W}_{\parallel}$
functional now is given by,
\begin{eqnarray}
    \mathcal{W}_{\parallel}&=&-\frac{e^2}{2}\frac{R^2}{T\,a}\left\{\frac{8}{3\varsigma}+\sum_{n=1}^{\infty}(-1)^n\left[\frac{\varsigma}{\pi^2}\frac{1}{n^2}+\frac{4}{\varsigma}-\sqrt{\frac{\pi}{2}}\frac{8\pi}{n^3\varsigma^2}\left(n^4+\frac{\varsigma^4}{16\pi^4}\right)e^{-\frac{1}{2}\left(\frac{n\pi}{\varsigma}\right)^2}\operatorname{Erfi}(\sqrt{2}\frac{n\pi}{\varsigma})\right]\right\}\nonumber\\
                           &\simeq&-\frac{e^2}{2}\frac{R^2}{T\,a}\left\{\frac{8}{3\varsigma}-\frac{\varsigma^3}{\pi^4}\sum_{n=1}^{\infty}(-1)^n\frac{1}{n^4}+\cdots\right\}\,,\qquad\text{for $\varsigma\ll 1$}\,.
\end{eqnarray}
Then, as $ a/T \gg 1$, we have
\begin{equation}
    \mathcal{W}_{\parallel}\simeq\mathcal{W}_0\left\{1+\frac{7\pi^4}{120}\,\frac{T^4}{a^4}+\cdots\right\}\,.
\end{equation}
The first term of the decoherence functional comes from the
influence of  vacuum fluctuations without the  boundary, while the
second term arises from the presence of the double-plate.

Some remarks are in place. In the perpendicular configuration, the
correction of the decoherence effect in Eq.~\eqref{pq} takes the
exponential form for the large $a/T$. This is due to the fact that
the double plate geometry provides a length scale $2a$, the plate
separation, in the $z$ direction, thus introducing this scale into
the $z$--$z$ components of the correlation functions. However, in
the parallel configuration, there is no such a length scale in this
direction. Thus, it ends up with the correction of the form of the
power of the ratio $a/T$ in the above expression. In addition, for
the result of the $\mathcal{W}_{\parallel}$ in  either  the
single-plate or the double-plate case, as shown in
Figs.~\ref{Fi:paraperp1} and \ref{Fi:paraperp2} respectively, we
observe that $\mathcal{W}_{\parallel}\simeq0$ for small  $ z_0/T$ or
$a/T$, and then $\mathcal{W}_{\parallel}$ approaches the value of
$\mathcal{W}_{0}$ from below in the region of large  $ z_0/T$ or
$a/T$. Thus, $\mathcal{W}_{\parallel}$ must intersect with
$\mathcal{W}_{0}$ at some value of the ratio, and exist a local
minimum in these cases.

\section{Brief discussion on finite conductivity effect}\label{sec4}
The finite conductivity effect on electron coherence due to electromagnetic fields near conducting plates have been discussed~\cite{AN}. In the interference experiment, when the electron moves parallel to the surface of the conducting plate with velocity $\mathbf{v}$, the induced surface charge in the conductor is expected to move along with the electron with the same velocity. As a result, the electric field inside the conductor in the direction of motion of the electron arises, and is to be $\mathbf{E}\propto e\,\mathbf{v}$. The induced current then is given by the Ohmic law, $\mathbf{J}=\sigma\mathbf{E}$, where $\sigma$ is the conductivity of the conductor. The presence of the current inside the conductor leads to energy loss due to Joule heating at a rate $ P_{\rm{Joule}}$, roughly given by $ P_{\mathrm{ Joule}}\propto\mathbf{E}\cdot\mathbf{J}\propto(e^2/\sigma)\,v^2$ with $v=\left|\mathbf{v}\right|$. We assume that the electron moves at a distance $d$ from the surface of the conducting plate. Then the $d$--dependence of the Joule energy loss rate $P_{\mathrm{Joule}}$ is given in Ref.~\cite{BO} in the context of classical electrodynamics,
\begin{equation}
P_{\mathrm{Joule}}=\frac{1}{16\pi}\biggl(\frac{e^2}{d}\biggr)\biggl(\frac{v^2}{\sigma d^2} \biggr)\,.
\end{equation}
For a resistive plate boundary at room temperature, the effect of Joule heating is found to play a key role on electron coherence in the interference experiment~\cite{AN}. The
observed contrast of electron interference fringes decreases due to large energy loss from Ohmic resistance as the electron moves close to the boundary. However, that is a different channel of decoherence from what we study. Here we consider electron decoherence due to vacuum fluctuations of electromagnetic fields with the perfectly conducting plate boundary. In contrast, when the electron travels parallel to the conducting plate, electron coherence is enhanced instead as it gets closer to the boundary since the electric field, responsible for decoherence, is vanishing along the plate surface.  It is
of interest to estimate the value of conductivity $\sigma$ at which the decoherence dynamics due to vacuum fluctuations is not masked by Ohmic dissipation. In the typical interference experiment, the electron moves with low velocities, and its velocity change is determined by the electric field with an overall factor $e/m$. The mean squared velocity dispersion owing to vacuum fluctuations of the fields along the surface of the conducting plate is given by~\cite{YU},
\begin{equation}
\langle\Delta v^2\rangle=\frac{1}{4\pi^2}\biggl(\frac{e^2}{d}\biggr)\biggl(\frac{1}{m\,d}\biggr)\biggl(\frac{1}{m}\biggr)\,,
\end{equation}
where the parameter $d$ is the distance of the electron to the plate. As long as
the Joule energy loss during the electron's flight time $T$ is much smaller than average energy fluctuations obtained from the velocity fluctuations above, the effect from the finite conductivity of the boundary can be ignored for the large enough conductivity given by:
\begin{equation}
\sigma\approx\frac{\pi}{2}\frac{mvL}{d}\approx 10^{20}\biggl(\frac{v}{10^{-4}c}\biggr) \biggl(\frac{10\,\mu\mathrm{m}}{d}\biggr)\biggl(\frac{L}{10\,\mathrm{cm}}\biggr)\mathrm{s}^{-1}\,,
\end{equation}
with the electron's path length $L=v\,T$. This required high conductivity roughly about two orders of magnitude larger than that of Copper at room temperature can possibly be achieved for metallic material at low temperature.

\section{Summary and Conclusion}\label{sec5}
In the present work, we investigate the influence of  zero-point
fluctuations of  quantum electromagnetic fields in the presence of
the perfectly conducting plates on  electrons. The effects of
modified vacuum fluctuations can be observed through the electron
interference experiment, and are manifested in the form of the
amplitude change and  phase shift of the interference fringes. Here
we first of all outline the closed-time-path formalism to describe
the evolution of the density matrix of the
 electron and fields.
Then, the  method of influence functional is employed
 by tracing out the fields
in the Coulomb gauge from which we find the evolution of the reduced
density matrix  of the electron with self-consistent backreaction.

Under the classical approximation with the prescribed electron's
trajectory dictated by an external potential, we find that the
exponent of the modulus of the influence functional describes the
extent of the amplitude change of the interference contrast, and its
phase results in an overall shift for the interference pattern. In
addition, it is known that the semiclassical Langevin equation for
considering the stochastic behavior of the particle coupled to
quantum  fields, involves backreaction dissipation in terms of the
retarded Green's function of fields as well as the accompanying
stochastic noise with the noise correlation function given by its
Hadamard function. These two effects are in general linked by the
fluctuation-dissipation theorem~\cite{WU2}. Thus, we may conclude
that reduction of coherence is driven by field fluctuations while
the phase shift results from backreaction dissipation through
particle creation  that influences the mean trajectory of the
electron.

We evaluate the decoherence functional of the electrons with the
boundary on quantum electromagnetic fields. The boundary conditions
can be imposed by the presence of either a single plate or double
parallel plates. In each case, the path plane on which the electrons
travel for the interference experiment can be parallel or
perpendicular to the plate(s). It is found that the effects of
coherence reduction of the electrons by zero-point fluctuations with
the boundary are strikingly deviated from that without the boundary.
Thus, the presence of the conducting plate anisotropically modifies
electromagnetic vacuum fluctuations that in turn influence the
decoherence dynamics of the electrons. In particular, as the
electrons are close to the plate, electron coherence is enhanced in
the case where the path plane of the electrons is parallel to the
plate. It is resulted from
 the suppression of zero-point fluctuations due to the boundary in the direction transverse to the plate. On
the other hand,  the electron coherence is reduced in the
perpendicular configuration where zero-point fluctuations are
boosted instead  along the direction longitudinal to the plate. In
addition, in the presence of double parallel plates boundary,
zero-point fluctuations seems to make the electrons more coherent in
the parallel configuration, but less coherent in the perpendicular
one, as compared with the single-plate boundary.

Thus, the loss of decoherence of the electrons can be understood
from zero-point fluctuations of electromagnetic fields given by the
Hadamard function of  vector potentials. On the other hand, the
backreaction dissipation through photon emission can influence the
mean trajectory of the electron, and in turn leads to the phase
shift on the electron inference pattern through the retarded Green's
function. We wish in our future work to address the issue of the
relation  between the amplitude change and phase shift of
interference fringes via the fluctuation-dissipation theorem, which
might be testable in the  interference experiment.

\begin{acknowledgments}
We would like to thank Chun-Hsien Wu, Bei-Lok Hu and Larry H. Ford
for stimulating discussions. We also thank Hing-Tong Cho for drawing
the attention to the reference~\cite{BN}. This work was supported in
part by the National Science Council, R. O. C. under grant
NSC93-2112-M-259-007.
\end{acknowledgments}

\begin{appendix}
\section{the decoherence functional in the Feynman gauge}\label{S:app1}
The decoherence functional $\mathcal{W}$ obtained in the Coulomb
gauge can be cast into the gauge invariant expression
\eqref{w-fun-E-B field}. Here, we illustrate the nature of the gauge
invariance by explicitly computing the decoherence functional with
an alternative gauge fixing. We choose the Feynman gauge as an
example, and then the Green's functions of the vector potentials in
the presence of the conducting plates can be obtained by the method
of the image charge~\cite{BR}.

In the following discussion, we assume that path function $\zeta(t)$
is required to be sufficiently smooth and an even function of
time $t$. The range of time $t$ extends from $-\infty$ to $+\infty$ such
that, for the motion of the electron to be physically meaningful,
the first time derivative of the path function must vanish at
endpoints, that is, $\dot{\zeta}(-\infty)=\dot{\zeta}(+\infty)=0$ in
this case.

\subsection{the single plate}
Consider a conducting plate lying at the $z=0$ plane. The
$\mathcal{W}$ functional, with the help of the image method, is
given by
\begin{eqnarray}
    \mathcal{W}&=&-\frac{e^2}{2}\oint_{\mathcal{C}}dx_{\mu}\oint_{\mathcal{C}}dx'_{\nu}\frac{1}{4\pi^2} \left[\frac{\eta^{\mu\nu}}{\triangle t^2-\triangle x_{\parallel}^2-(z-z')^2}-\frac{\left(\eta^{\mu\nu}+2n^{\mu}n^{\nu}\right)}{\triangle t^2-\triangle x_{\parallel}^2-(z+z')^2}\right]\nonumber\\
   &=&\mathcal{W}^{(0)}+\mathcal{W}^{(R)}\,,
\end{eqnarray}
where $n^{\mu}=(0,0,0,1)$ is a unit vector normal to the plate.
$\triangle t$ and $\triangle x_{\parallel}$ denote $t-t'$ and
$x-x'$, respectively. Apparently, the $\mathcal{W}$ functional can
be written as the sum of $\mathcal{W}^{(0)}$ from the vacuum
fluctuations in the unbounded space and $\mathcal{W}^{(R)}$ from the
contribution of the image charge that accounts for the presence of
the conducting plate. The $\mathcal{W}^{(0)}$ term is explicitly
given by,
\begin{eqnarray}
    \mathcal{W}^{(0)}&=&-\frac{e^2}{8\pi^2}\,\oint_{\mathcal{C}}dx_{\mu}\oint_{\mathcal{C}}dx'_{\nu}\;\frac{\eta^{\mu\nu}}{\triangle t^2-\triangle x_{\parallel}^2-(z-z')^2}\nonumber\\
                 &=&\frac{e^2}{4}\left[\int_{\mathcal{C}_1}\int_{\mathcal{C}_1}+\int_{\mathcal{C}_2}\int_{\mathcal{C}_2}-\int_{\mathcal{C}_1}\int_{\mathcal{C}_2}-\int_{\mathcal{C}_2}\int_{\mathcal{C}_1}
\right]dt\,dt'\;( 1-\mathbf{v}\cdot\mathbf{v}')\nonumber\\
                 && \qquad\qquad\times\int\!\frac{d^3\mathbf{k}}{(2\pi)^3}\;\frac{1}{2\omega}\Bigl[e^{i\mathbf{k}^{\vphantom{'}}_{\parallel}\cdot(\mathbf{x}^{\vphantom{'}}_{\parallel}-\mathbf{x}'_{\parallel})+ik_z(z-z')-i\omega(t-t')}+\text{C.C.}\Bigr]\,,
\end{eqnarray}
where $\mathbf{v}=d\mathbf{x}/dt$ and
$\mathcal{C}=\mathcal{C}_1-\mathcal{C}_2$. We only consider the case
that the path plane is perpendicular to the plate and denote the
decoherence functional  as $\mathcal{W}_{\perp}$. The extension to
the parallel case is straightforward.

The worldlines of the electrons are chosen to take the form,
$\mathcal{C}_{1,2}=(t,v_xt,0,z_0\pm\zeta(t))$. The Lorentz
invariance of the decoherence functional enables us to choose the
frame $\mathfrak{S}$ moving along a straight line described by
$(t,v_xt,0,z_0)$ defined in Sec.~\ref{sec4}. Observed from this
reference frame, the electrons are to move only transversally in the
$z$ direction. Then, the $ \mathcal{W}_{\perp}^{(0)}$ term reduces
to
\begin{align}
   \mathcal{W}_{\perp}^{(0)}&=e^2\left\{\int\!dt\,dt'\;(1-\dot{\zeta}\dot{\zeta}')\int\!\frac{d^3\mathbf{k}}{(2\pi)^3}\;\frac{1}{2\omega}\;e^{-ik_z(\zeta-\zeta')+i\omega(t-t')}\right.\notag\\
                    &\qquad\left.\qquad\qquad\qquad-\int\!dt\,dt'\;(1+\dot{\zeta}\dot{\zeta}')\int\!\frac{d^3\mathbf{k}}{(2\pi)^3}\;\frac{1}{2\omega}\;e^{-ik_z(\zeta+\zeta')+i\omega(t-t')}\right\}\\
                    &=2e^2\left\{\int\!dt\,dt'\int\!\frac{d^3\mathbf{k}}{(2\pi)^3}\;\frac{1}{2\omega}\,\sin k_z\zeta\,\sin k_z\zeta'\,e^{i\omega(t-t')}\right.\nonumber\\
                    &\qquad\qquad\qquad\qquad\left.-\int\!dt\,dt'\;\dot{\zeta}\dot{\zeta}'\int\!\!\frac{d^3\mathbf{k}}{(2\pi)^3}\;\frac{1}{2\omega}\,\cos k_z\zeta\,\cos k_z\zeta'\,e^{i\omega(t-t')}\right\}\,.\label{jf}
\end{align}
We then perform the integration by parts on the first term of
Eq.~\eqref{jf} and obtain
\begin{equation}
   \mathcal{W}_{\perp}^{(0)}=-2e^2\int\!\frac{d^3\mathbf{k}}{(2\pi)^3}\;\frac{1}{2\omega}\left[1-\frac{k_z^2}{\omega^2}\right]\left|\int\!dt\;\dot{\zeta}\cos k_z\zeta\,e^{i\omega t}\right|^2\,.\label{rt}
\end{equation}
Following the similar procedures
leads to the $\mathcal{W}_{\perp}^{(R)}$ functional given by,
\begin{align}
   \mathcal{W}_{\perp}^{(R)}&=\frac{e^2}{8\pi^2}\oint_{\mathcal{C}}\!dx_{\mu}\oint_{\mathcal{C}}\!dx_{\nu}\sum_{n=-\infty}^{\infty}\frac{\eta^{\mu\nu}-n^{\mu}n^{\nu}}{\triangle t^2-\triangle x_{\parallel}^2-(z-z'-2na)^2}\notag\\
                             &=-2e^2\int\!\frac{d^3\mathbf{k}}{(2\pi)^3}\;\frac{1}{2\omega}\,e^{-2ik_zz_0}\left[1-\frac{k_z^2}{\omega^2}\right]\left|\int\!dt\;\dot{\zeta}\cos k_z\zeta\,e^{i\omega t}\right|^2\,.\label{th}
\end{align}
Then, putting Eqs.~\eqref{rt} and~\eqref{th} together, the
$\mathcal{W}_{\perp}$ functional becomes
\begin{equation}
   \mathcal{W}_{\perp}=-2e^2\int\!\frac{d^3\mathbf{k}}{(2\pi)^3}\;\frac{1}{2\omega}\biggl[1+e^{-2ik_zz_0}\biggr]\biggl[ 1-\frac{k_z^2}{\omega^2}\biggr] \left|\int\!dt\;\dot{\zeta}\cos(k_z\,\zeta)e^{i\omega t}\right|^2\,,
\end{equation}
which is of the same form as Eq.~\eqref{W_perp_single} derived in
the Coulomb gauge.

\subsection{the double-plate}
We now turn to the case with two conducting plates at the $z=0$ and
$z=a$ planes, respectively. The $\mathcal{W}$ functional is given by
\begin{eqnarray}
    \mathcal{W}&=&-\frac{e^2}{8\pi^2}\oint_{\mathcal{C}}dx_{\mu}\oint_{\mathcal{C}}dx'_{\nu}\sum_{n=-\infty}^{\infty}\left[\frac{\eta^{\mu\nu}}{\triangle t^2-\triangle x_{\parallel}^2-(z-z'-2na)^2}-\frac{\eta^{\mu\nu}+2n^{\mu}n^{\nu}}{\triangle t^2-\triangle x_{\parallel}^2-(z+z'-2na)^2}\right]\nonumber\\
               &=&\mathcal{W}^{(I)}+\mathcal{W}^{(I\!I)}\,,\label{w-normal-double-feyn}
\end{eqnarray}
in terms of a sum of the contributions from the image charges. The
$\mathcal{W}^{(I)}$ can be written explicitly as
\begin{align}
    \mathcal{W}^{(I)}&=-\frac{e^2}{8\pi^2}\oint_{\mathcal{C}}\!dx_{\mu}\oint_{\mathcal{C}}\!dx_{\nu}\sum_{n=-\infty}^{\infty}\frac{\eta^{\mu\nu}}{\triangle t^2-\triangle x_{\parallel}^2-(z-z'-2na)^2}\notag\\
                           &=\frac{e^2}{2}\left[\int_{\mathcal{C}_1}\int_{\mathcal{C}_1}+\int_{\mathcal{C}_2}\int_{\mathcal{C}_2}-\int_{\mathcal{C}_1}\int_{\mathcal{C}_2}-\int_{\mathcal{C}_2}\int_{\mathcal{C}_1}
\right]dt\,dt'\;( 1-\mathbf{v}\cdot\mathbf{v}')\\
                           &\qquad\qquad\qquad\times\sum_{n=-\infty}^{\infty}\int\frac{d^3\mathbf{k}}{(2\pi)^3}\;\frac{1}{2\omega}\left[e^{i\mathbf{k}^{\vphantom{'}}_{\parallel}\cdot(\mathbf{x}^{\vphantom{'}}_{\parallel}-\mathbf{x}'_{\parallel})+ik_z(z-z'-2na)-i\omega(t-t')}+\text{C.C.}\right]\,,\notag
\end{align}
with the velocity $\mathbf{v}=d\mathbf{x}/dt$ and the closed path
given by $\mathcal{C}=\mathcal{C}_1-\mathcal{C}_2$.
The integration over $k_z$ can be carried out by the identity
\begin{equation}
    \sum_{n=-\infty}^{\infty}e^{2ik_zna}=\frac{\pi}{a}\sum_{m=-\infty}^{\infty}\delta(k_z-\frac{m\pi}{a})\,.
\end{equation}
We consider the case that the path plane of the electrons is
perpendicular to the plates and the worldlines of electrons are
described by $\mathcal{C}_{1,2}=(t,v_xt,0,\frac{a}{2}\pm\zeta(t))$.
We  evaluate the decoherence functional $\mathcal{W}_{\perp}$ in the
frame $\mathfrak{S}$ with $z_0=a/2$. Therefore, the
$\mathcal{W}_{\perp}$ function is simplified to
\begin{eqnarray}
   \mathcal{W}_{\perp}^{(I)}&=&\frac{e^2}{2a}\sum_{n=-\infty}^{\infty}\left\{\int\!dtdt'\;(1-\dot{\zeta}\dot{\zeta}')\int\!\frac{d^2\mathbf{k}_{\parallel}}{(2\pi)^2}\;\frac{1}{2\omega_n}\,e^{-i\frac{n\pi}{a}(\zeta-\zeta')+i\omega_n(t-t')}\right.\nonumber\\
                     &&\qquad\qquad\left.-\int\!dtdt'\;(1+\dot{\zeta}\dot{\zeta}')\int\!\frac{d^2\mathbf{k}_{\parallel}}{(2\pi)^2}\;\frac{1}{2\omega_n}\,e^{-i\frac{n\pi}{a}(\zeta+\zeta')+i\omega_n(t-t')}\right\}\nonumber\\
                     &=&\frac{e^2}{a}\sum_{n=-\infty}^{\infty}\left\{\int\!dtdt'\int\!\frac{d^2\mathbf{k}_{\parallel}}{(2\pi)^2}\;\frac{1}{2\omega_n}\sin\frac{n\pi}{a}\zeta\,\sin\frac{n\pi}{a}\zeta'\;e^{i\omega_n(t-t')}\right.\nonumber\\
                     &&\qquad\qquad\left.-\int\!dtdt'\;\dot{\zeta}\dot{\zeta}'\int\!\frac{d^2\mathbf{k}_{\parallel}}{(2\pi)^2}\;\frac{1}{2\omega_n}\cos\frac{n\pi}{a}\zeta\,\cos\frac{n\pi}{a}\zeta'\;e^{i\omega_n(t-t')}\right\}\,.
\end{eqnarray}
Taking the integration by parts for the first term of the above
expression, the $\mathcal{W}_{\perp}^{(I)}$ functional ends up with
\begin{equation}
   \mathcal{W}_{\perp}^{(I)}=-\frac{e^2}{a}\sum_{n=-\infty}^{\infty}\int\!\!\frac{d^2\mathbf{k}_{\parallel}}{(2\pi)^2}\;\frac{1}{2\omega_n}\left[1-\frac{n^2 \pi^2 }{\omega_n^2 a^2 }\right]\left|\int\!dt\;\dot{\zeta}\cos\frac{n\pi}{a}\zeta\;e^{i\omega_nt}\right|^2\,.
\end{equation}
Following the similar procedures, we come to
\begin{eqnarray}
    \mathcal{W}_{\perp}^{(I\!I)}&=&\frac{e^2}{8\pi^2}\oint_{\mathcal{C}}dx_{\mu}\oint_{\mathcal{C}}dx'_{\nu}\sum_{n=-\infty}^{\infty}\frac{\eta^{\mu\nu}+2n^{\mu}n^{\nu}}{\triangle t^2-\triangle x_{\parallel}^2-(z+z'-2na)^2}\nonumber\\
                                  &=&-\frac{e^2}{a}\sum_{n=-\infty}^{\infty}\int\!\frac{d^2\mathbf{k}_{\parallel}}{(2\pi)^2}\;\frac{(-1)^n}{2\omega_n}\left[1-\frac{n^2 \pi^2 }{\omega_n^2 a^2 }\right]\left|\int\!dt\;\dot{\zeta}\cos\frac{n\pi}{a}\zeta\;e^{i\omega_nt}\right|^2\,.
\end{eqnarray}
Then, the sum of two contributions give rise to the $\mathcal{W}_{\perp}$
function of the form:
\begin{equation}
    \mathcal{W}_{\perp}=-\frac{2e^2}{a}\sum_{n=\text{even}}\int\!\frac{d^2\mathbf{k}_{\parallel}}{(2\pi)^2}\;\frac{1}{2\omega_n}\left[1-\frac{n^2 \pi^2 }{\omega_n^2 a^2 }\right]\left|\int\!dt\;\dot{\zeta}\cos\frac{n\pi}{a}\zeta\;e^{i\omega_nt}\right|^2\,,
\end{equation}
which is also consistent with the result Eq~\eqref{W-normal-double} in the Coulomb gauge.

\section{transformation of the slowly convergent summation}\label{S:app2}
Here we  outline the  method to convert an expression of summation,
which turns out to be slowly convergent, into another form to carry
out the sum much efficiently~\cite{BN}. In general, one may express
a summation by a contour integral,
\begin{align}
    \sum_{n=1}^{\infty}f(n)&=\oint_{\Gamma}dz\;\frac{f(z)}{e^{2i\pi z}-1}\notag\\
                           &=\left\{\int_{\Gamma_1}+\int_{\Gamma_2}+\int_{\Gamma_3}+\int_{\Gamma_4}\right\}dz\;\frac{f(z)}{e^{2i\pi z}-1}\,,
\end{align}
where the closed path $\Gamma$ is chosen to enclose all simple poles
at $z\in\mathbb{Z}^+$ in a counterclockwise sense, and otherwise
quite arbitrary. It proves convenient to express the closed contour
$\Gamma$ with the following 4 segments,
\begin{align*}
    \Gamma_1&:z=s-i\epsilon\,,& \delta<&s<\infty\,,\\
    \Gamma_2&:z=\infty+is\,,& -\epsilon<&s<\epsilon\,,\\
    \Gamma_3&:z=s+i\epsilon\,,&\delta<&s<\infty\,,\\
    \Gamma_4&:z=\delta+is\,,& -\epsilon<&s<\epsilon\,,
\end{align*}
where $\epsilon\to0^+$ and $0\leq\delta<1$. The contour integral can
be carried out as follows.

Since $\Gamma_1$ lies just below the real axis, one may expand the
denominator of the integrand in terms of $e^{-2i\pi z}$ to ensure
convergence of the integral
\begin{align}
    \int_{\Gamma_1}dz\;\frac{f(z)}{e^{2i\pi z}-1}&=\int_{\delta-i\epsilon}^{\infty-i\epsilon}\!ds\;\frac{f(s)}{e^{2i\pi s}-1}\notag\\
                                            &=\sum_{k=1}^{\infty}\int_{\delta}^{\infty}\!ds\;f(s)\,e^{-2i\pi ks}\,.
\end{align}
On the contrary,  the denominator of the integrand is expanded with
respect to $e^{2i\pi z}$ as the path $\Gamma_3$ lies slightly above
the real axis as
\begin{align}
    \int_{\Gamma_3}dz\,\frac{f(z)}{e^{2i\pi z}-1}&=-\int_{\delta+i\epsilon}^{\infty+i\epsilon}\!ds\;\frac{f(s)}{e^{2i\pi s}-1}\notag\\
                                            &=\sum_{k=0}^{\infty}\int_{\delta}^{\infty}\!ds\;f(s)\,e^{2i\pi ks}\notag\\
                                            &=\int_{\delta}^{\infty}ds\;f(s)+\sum_{k=1}^{\infty}\int_{\delta}^{\infty}\!ds\;f(s)\,e^{2i\pi ks}\,.
\end{align}
The line integral along the path $\Gamma_2$ turns out to be zero
\begin{equation}
    \int_{\Gamma_2}dz\;\frac{f(z)}{e^{2i\pi z}-1}=\int_{\infty-i\epsilon}^{\infty+i\epsilon}ds\;\frac{f(s)}{e^{2i\pi s}-1}=0
\end{equation}
for a regular function $f(z)$. Special care must be taken for the
line integral along the path $\Gamma_4$. The value of $\delta$ can
be chosen within $0\leq\delta<1$,  which leads to the same result of
the contour integral. However, in the limit $\delta\to 0$, the path
may come across a pole at $z=0$ so it must be deformed to avoid the
pole. Then, in this case, the path $\Gamma_4$ can be chosen to be a
semicircle connecting $0+i\epsilon$ and $0-i\epsilon$ clockwise,
that is, $z=\epsilon e^{i\theta}$ with $-\pi/2<\theta<\pi/2$. The
line integral then becomes
\begin{align}
    \int_{\Gamma_4}dz\,\frac{f(z)}{e^{2i\pi z}-1}&=-i\epsilon\int_{-\frac{\pi}{2}}^{\frac{\pi}{2}}d\theta\;\frac{e^{i\theta}}{e^{2i\pi\epsilon^{i\theta}}-1}\,f(\epsilon\,e^{i\theta})=-\frac{1}{2}f(0)\,.
\end{align}
However, for a non-zero $\delta$, the integral vanishes just as that
over the path $\Gamma_2$,
\begin{equation}
    \int_{\Gamma_4}dz\;\frac{f(z)}{e^{2i\pi z}-1}=-\int_{\delta-i\epsilon}^{\delta+i\epsilon}ds\;\frac{f(s)}{e^{2i\pi s}-1}=0\,.
\end{equation}
Putting these results together, we have
\begin{equation}\label{E:ft}
    \sum_{n=1}^{\infty}f(n)=\oint_{\Gamma}dz\,\frac{f(z)}{e^{2i\pi z}-1}=-\frac{1}{2}f(0)\,\delta_{\delta0}+\int_{\delta}^{\infty}ds\,f(s)+2\sum_{k=1}^{\infty}\int_{\delta}^{\infty}ds\,f(s)\cos(2i\pi ks)\,.
\end{equation}
The third term in the right hand side of the above repression is
essentially a Fourier transformation of $f(s)$, which  transforms
the variable $s$ to variable $k$ roughly related by $k\approx1/s$.
Thus,  when the summation $\sum_nf(n)$ converges slowly,  the
summation shown in the right hand side of Eq.~\eqref{E:ft} will be
carried out much efficiently.
\end{appendix}


\begin{thebibliography}{99}
\bibitem{ZU}
W. H. Zurek, Phys. Rev. D {\bf 24}, 1516 (1981); {\it ibid} {\bf
26}, 1862 (1982); Phys. Today {\bf 44}, 36 (1991); W. G. Unruh and
W. H. Zurek, Phys. Rev. D {\bf 40}, 1071 (1989); J. P. Paz, S. Habib
and W. H. Zurek, Phys. Rev. D {\bf 47}, 488 (1993); W. H. Zurek, S.
Habib and J. P. Paz, Phys. Rev. Lett. {\bf 70}, 1187 (1993); W. H.
Zurek, Prog. Theor. Phys. {\bf 89}, 281 (1993).


\bibitem{HA}
M. Gell-Mann and J. B. Hartle, Phys. Rev. D {\bf 47}, 3345 (1993);
J. B. Hartle, in {\it Directions in General Relativity}, Vol. 1,
edited by B. L. Hu, M. P. Ryan, and C. V. Vishveswara (Cambridge
University Press, 1993); H. F. Dowker and J. J. Halliwell, Phys.
Rev. D {\bf 46}, 1580 (1992).

\bibitem{CAL1}
E. Calzetta and B. L. Hu, in {\it Directions in General Relativity},
edited by B. L. Hu and T. A. Jacobson (Cambridge University Press,
1993), Vol. II; E. Calzetta and B. L. Hu, in {\it Heat Kernel
Techniques and Quantum Gravity}, edited by S. A. Fulling (Texas A \&
M Press, 1995).

\bibitem{CAL2}
E. Calzetta and B. L. Hu, Phys. Rev. D {\bf 52},  6770 (1995).

\bibitem{CAD}
A. O. Caldeira and A. J. Leggett, Phys. Rev. A {\bf 31}, 1059
(1985); P. M. V. B. Barone and A. O. Caldeira, Phys. Rev. A {\bf
43}, 57 (1991); B. L. Hu, J. P. Paz, and Y. Zhang, Phys. Rev D {\bf
45}, 2843 (1992); B. L. Hu, J. P. Paz, and Y. Zhang, Phys. Rev D
{\bf 47}, 1576 (1993); J. R. Anglin, J. P. Paz, and W. H. Zurek,
Phys. Rev. A {\bf 55}, 4041 (1997).

\bibitem{ST}
A. A. Starobinsky, in {\it Field Theory, Quantum Gravity and Strings}, edited by H. J. de Vega and N. Sanchez (Springer, Berlin, 1986).

\bibitem{RE}
S. J. Rey, Nucl. Phys. B {\bf 284}, 706 (1987); R. Brandenberger, R.
Laflamme, and M. Mijic, Mod. Phys. Lett. A {\bf 5}, 2311 (1990); W.
Lee, Y.-Y. Charng, D.-S. Lee, and L.-Z. Fang, Phys. Rev. D {\bf 69},
123522 (2004).

\bibitem{NI}
M. Nielsen and I. Chuang, {\it Quantum Computation and Quantum Information}, (Cambridge University Press, 2000).

\bibitem{FO}
L. H. Ford, Phys. Rev. D {\bf 47}, 5571 (1993); Phys. Rev. A {\bf 56}, 1812 (1997).

\bibitem{MA}
F. D. Mazzitelli, J. P. Paz, and A. Villanueva, Phys. Rev. A {\bf 68}, 062106 (2003).

\bibitem{CA}
H. B. G. Casimir, Proc. K. Ned. Akad. Wet. {\bf 51}, 793 (1948).

\bibitem{BA1}
G. Barton, J. Phys. A {\bf 24}, 991 (1991); {\it ibid} 5533 (1991).

\bibitem{BA2}
G. Barton,  Proc. Roy. Soc. Lond. A {\bf 320}, 251 (1970).

\bibitem{RO}
D. Robaschik and E. Wieczorek, Ann. Phys. (N.Y.) {\bf 236}, 43 (1994);
{\it ibid}  Phys. Rev. D {\bf 52}, 2341 (1995).

\bibitem{SC}
J. Schwinger, J. Math. Phys. {\bf 2}, 407 (1961); L. V. Keldysh,
Sov. Phys. JETP {\bf 20}, 1018 (1965); K. T. Mahanthappa, Phys. Rev.
{\bf 126}, 329 (1962); P. M. Bakshi and K. T. Mahanthappa, J. Math.
Phys. {\bf 4}, 1 (1963); {\it ibid.} {\bf 4}, 12 (1963); K.-C. Chou,
Z.-B. Su, B.-L. Hao, and L. Yu, Phys. Rep. {\bf 118}, 1 (1985); J.
Rammer and H. Smith, Rev. Mod. Phys. {\bf 58}, 323 (1986).

\bibitem{LEE}
Y.-Y. Charng, K.-W. Ng, C.-Y. Lin, and D.-S. Lee, Phys. Lett. B {\bf
548},175 (2002); Y.-Y. Charng, D.-S. Lee, C. N. Leung, and K.-W. Ng,
hep-ph/0506273.

\bibitem{BR}
L. S. Brown and G. J. Maclay, Phys. Rev. {\bf 184}, 1272 (1969).

\bibitem{HS}
J.-T. Hsiang and L. H. Ford, Phys. Rev. Lett. {\bf 92}, 250402
(2004).

\bibitem{WU2}
C.-H. Wu and D.-S. Lee,  Phys. Rev. D {\bf 71}, 125005 (2005).

\bibitem{YU}
H. Yu and L. H. Ford, Phys. Rev. D {\bf 70}, 065009 (2004).

\bibitem{BN}
H. Boschi-Filho, C. P. Natividade and C. Farina, Phys. Rev. D {\bf
45}, 586 (1992).

\bibitem{AN}
J. R. Anglin and W. H. Zurek, in {\it Dark Matter in Cosmology,
Quantum Measurements, Experimental Gravitation}, Proc. XXXIst
Rencontres de Moriond, edited by R. Ansari, Y. Giraud-H\'{e}raud,
and J. Tr\^{a}n Thanh V\^{a}n (Editions Frontieres, Gif-sur-Yvette,
1996). quant-ph/9611049; Y. Levinson, J. Phys. A {\bf 37}, 3003
(2004); P. Sonnentag and F. Hasselbach, Braz. J. Phys. {\bf 35}, 385
(2005).

\bibitem{BO} T. H. Boyer, Phys. Rev. A {\bf 9}, 68 (1974).
\end{thebibliography}
\end{document}